\documentclass[twocolumn,eqsecnum,aps]{revtex4}
\usepackage{amsmath,amsfonts,amsbsy,amssymb}
\usepackage[mathscr]{euscript}
\begin{document}

\title{Noncommutative Geometry, Extended W$_{\infty}$ Algebra and
Grassmannian Solitons\\
in Multicomponent Quantum Hall Systems }
\author{Z.F. Ezawa$^1$, G. Tsitsishvili$^2$ and K. Hasebe$^3$}
\affiliation{{}$^1$Department of Physics, Tohoku University, Sendai, 980-8578 Japan }
\affiliation{{}$^2$Department of Theoretical Physics, A. Razmadze Mathematical Institute,
Tbilisi, 380093 Georgia}
\affiliation{{}$^3$Department of Physics, Sogang University, Seoul 100-611, Korea}

\begin{abstract}
Noncommutative geometry governs the physics of quantum Hall (QH) effects. We
introduce the Weyl ordering of the second quantized density operator to
explore the dynamics of electrons in the lowest Landau level. We analyze QH
systems made of $N$-component electrons at the integer filling factor $%
\nu=k\leq N$. The basic algebra is the SU(N)-extended W$_{\infty}$. A
specific feature is that noncommutative geometry leads to a spontaneous
development of SU(N) quantum coherence by generating the exchange Coulomb
interaction. The effective Hamiltonian is the Grassmannian $G_{N,k}$ sigma
model, and the dynamical field is the Grassmannian $G_{N,k}$ field,
describing $k(N-k)$ complex Goldstone modes and one kind of topological
solitons (Grassmannian solitons).
\end{abstract}

\maketitle


\section{Introduction}

Noncommutative geometry\cite{BookConnes} has found growing attention in
field theory and superstring theory\cite{Connes98,Douglas98,Seiberg99}.
However, its physical evidence is still very rare. The quantum Hall (QH)
effect provides a rare evidence\cite{BookEzawa}, where all physics results
from the noncommutative geometry in the plain, 
\begin{equation}
\lbrack X,Y]=-i\ell _{B}^{2}.  \label{NonCommuXY}
\end{equation}%
Here, $(X,Y)$ describes the position of the planar electron confined to the
lowest Landau level (LLL), and $\ell _{B}$ is the magnetic length proving
the scale. As has been discussed extensively\cite%
{Girvin86B,Iso92PLB,Cappelli93NPB}, the QH system is characterized by the W$%
_{\infty }$ algebra.

A new aspect of QH systems\cite{BookEzawa,BookDasSarma} is quantum coherence
due to the spin degree of freedom, which is also a consequence of the
noncommutativity (\ref{NonCommuXY}). Electron spins are polarized
spontaneously rather than compulsively by the Zeeman effect: Hence, the
system is called QH ferromagnet. The basic algebraic structure is the
SU(2)-extended W$_{\infty }$\cite{Ezawa97B}. Topological solitons ($CP^{1}$
solitons) arise as coherent excitations\cite{Sondhi93B}, which have been
observed experimentally\cite{Barrett95L,Aifer96L,Schmeller95L}. Much more
interesting phenomena occur in bilayer QH systems. For instance, an
anomalous tunneling current has been observed\cite{Spielman00L} between the
two layers at the zero bias voltage. It may well be a manifestation of the
Josephson-like phenomena predicted a decade ago\cite{Ezawa92IJMPB}. They
occur due to quantum coherence developed spontaneously across the layers\cite%
{Ezawa97B,Moon95B}. QH effects present experimental tests of various ideas
inherent to noncommutative geometry.

In this paper we investigate the algebraic structure of the $N$-component QH
system subject to the noncommutativity (\ref{NonCommuXY}). We then analyze
the spontaneous symmetry breaking at the filling factor $\nu =k$, $%
k=1,2,\ldots ,N$, and show that the Goldstone modes and topological solitons
are described by the Grassmannian $G_{N,k}$ field. Here, the $G_{N,k}$ field
is the one that takes values on the Grassmannian $G_{N,k}$ manifold. Note
that the $G_{N,1}$ manifold is equal to the $CP^{N-1}$ manifold. Such a
multicomponent QH system is feasible experimentally by constructing $%
N^{\prime }$-layer QH systems, where $N=2N^{\prime }$ with the spin degree
of freedom and $N=N^{\prime }$ without it. See Refs.\cite{Hasebe02B} for a
specific application to bilayer QH ferromagnets where $N=4$. As far as we
are aware of, this is the first established system where the Grassmannian $%
G_{N,k}$ field plays a key role in physics.

In Sections II and III we review the noncommutative planar system and the
LLL projection, respectively. We make a novel proposition of the Weyl
ordering of the second quantized density operator. In Section IV we derive
the SU(N)-extended W$_{\infty }$ as the algebraic structure of
multicomponent electrons in the noncommutative planar system. In Section V,
employing an algebraic method, we represent the Coulomb potential so that
the exchange interaction effect is made manifest. The exchange Coulomb
interaction is the key to quantum coherence. In Section VI, we make a
derivative expansion and derive the SU(N) nonlinear sigma model as an
effective Hamiltonian. It yields the Grassmannian $G_{N,k}$ sigma model for
the QH system at $\nu =k$. In Section VII we show that the dynamic field is
the $G_{N,k}$ field describing $k(N-k)$ complex Goldstone modes. In Section
VIII we construct Grassmannian $G_{N,k}$ solitons as topological objects. In
Section IX, by reexamining the LLL projection, we discuss what
quasiparticles we expect to observe in QH systems. In Section X we make a
brief application of our results to realistic QH systems by including the
Zeeman and tunneling interactions. Note that all Goldstone modes are made
massive due to these interactions. Section XI is devoted to discussions.

\section{Noncommutative Planar Systems}

\label{SecNonCommu}

The position of an electron confined to the lowest Landau level is specified
by the guiding center $\boldsymbol{X}=(X,Y)$ subject to the noncommutativity (%
\ref{NonCommuXY}). There exists a procedure (the Weyl prescription\cite%
{Weyl50,Wong98}) to construct a noncommutative theory with the coordinate $%
\boldsymbol{X}=(X,Y)$ from a commutative theory with the coordinate $\boldsymbol{x}%
=(x,y)$. From a function $f(\boldsymbol{x})$ in the commutative space, we
construct 
\begin{equation}
\widehat{O}_{f}=\frac{1}{(2\pi )^{2}}\int \!d^{2}qd^{2}x\,e^{-i\boldsymbol{q}(%
\boldsymbol{x}-\boldsymbol{X})}f(\boldsymbol{x}).  \label{WeylTrans}
\end{equation}%
Taking the plane wave $f(\boldsymbol{x})=e^{i\boldsymbol{px}}$ in (\ref{WeylTrans}),
we find%
\begin{equation}
\widehat{O}_{f}=e^{i\boldsymbol{pX}}.  \label{WeylOrder}
\end{equation}%
We call it the Weyl-ordered plane wave. It approaches the plane wave $e^{i%
\boldsymbol{px}}$ in the commutative limit ($\ell _{B}\rightarrow 0$).

It is convenient to construct a Fock representation of the algebra (\ref%
{NonCommuXY}) by way of the operators,%
\begin{equation}
b=\frac{1}{\sqrt{2}\ell _{B}}(X-iY),\qquad b^{\dag }=\frac{1}{\sqrt{2}\ell
_{B}}(X+iY),  \label{WeylXb}
\end{equation}%
obeying $[b,b^{\dag }]=1$. The Fock space is made of the states 
\begin{equation}
|n\rangle =\frac{1}{\sqrt{n!}}(b^{\dag })^{n}|0\rangle ,\qquad
n=0,1,2,\cdots .  \label{LandaSite}
\end{equation}%
They are the quantum mechanical states inherent to the noncommutativity (\ref%
{NonCommuXY}). We call them the Landau sites in QH systems.

Two of the Landau sites are related as%
\begin{equation}
|m\rangle =\sqrt{\frac{n!}{m!}}L(m,n)|n\rangle ,
\end{equation}%
with $L(m,n)=(b^{\dagger })^{m}b^{n}$. They generate the algebra, 
\begin{equation}
\left[ L(m,n),L(k,l)\right] =\sum_{t=1}^{\infty }C_{mn;kl}^{t}L(m+k-t,n+l-t),
\label{WinftyB}
\end{equation}%
where the structure constant is 
\begin{equation}
C_{mn;kl}^{t}={\frac{1}{t!}}\left( {\frac{n!k!}{(n-t)!(k-t)!}}-{\frac{m!l!}{%
(m-t)!(l-t)!}}\right) .
\end{equation}%
This is the W$_{\infty }$ algebra\cite{Iso92PLB,Cappelli93NPB}
characterizing the noncommutative planar system.

The noncommutative coordinate $\boldsymbol{X}=(X,Y)$ acts on the Landau site as%
\begin{align}
X|n\rangle & =\frac{\ell _{B}}{\sqrt{2}}\left[ \sqrt{n}|n-1\rangle +\sqrt{n+1%
}|n+1\rangle \right] ,  \notag \\
Y|n\rangle & =\frac{i\ell _{B}}{\sqrt{2}}\left[ \sqrt{n}|n-1\rangle -\sqrt{%
n+1}|n+1\rangle \right] .  \label{ActioBx}
\end{align}%
We may represent the algebra (\ref{NonCommuXY}) by the differential
operators,%
\begin{equation}
X(\boldsymbol{x})={\frac{1}{2}}x-i\ell _{B}^{2}{\frac{\partial }{\partial y}}%
,\qquad Y(\boldsymbol{x})={\frac{1}{2}}y+i\ell _{B}^{2}{\frac{\partial }{%
\partial x}}.  \label{RepreDiffe}
\end{equation}%
Then, $\widehat{O}_{f}$ acts on the Fock space via (\ref{ActioBx}), and it
is represented by a differential operator,%
\begin{equation}
\langle \boldsymbol{x}|\widehat{O}_{f}|n\rangle =\widehat{O}_{f}(\boldsymbol{x}%
)\langle \boldsymbol{x}|n\rangle ,
\end{equation}%
acting on the wave function. In this representation the wave function reads%
\begin{equation}
\mathfrak{S}_{n}(\boldsymbol{x})=\langle \boldsymbol{x}|n\rangle =\sqrt{\frac{1}{%
2^{n+1}\pi \ell _{B}^{2}n!}}\left( \frac{z}{\ell _{B}}\right) ^{n}e^{-%
\boldsymbol{x}^{2}/4\ell _{B}^{2}},  \label{OneBodyLLL}
\end{equation}%
with $z=x+iy$. It is seen from this wave function that each Landau site $%
|n\rangle $\ occupies area $2\pi \ell _{B}^{2}.$

The Weyl-ordered plane wave $e^{i\boldsymbol{pX}}$ generates the projective
translation group,%
\begin{equation}
e^{i\boldsymbol{pX}}e^{i\boldsymbol{qX}}=e^{i(\boldsymbol{p}+\boldsymbol{q})\boldsymbol{X}}\exp %
\left[ \frac{i}{2}\ell _{B}^{2}\boldsymbol{p}\!\wedge \!\boldsymbol{q}\right] ,
\label{MagneTrans}
\end{equation}%
with $\boldsymbol{p}\!\wedge \!\boldsymbol{q}=p_{x}q_{y}-p_{y}q_{x}$, as follows
from the noncommutativity (\ref{NonCommuXY}). We present two important
relations,%
\begin{equation}
\text{Tr}\left[ e^{i\boldsymbol{pX}}\right] \equiv \sum_{n=0}^{\infty }\langle
n|e^{i\boldsymbol{pX}}|n\rangle =\frac{2\pi }{\ell _{B}^{2}}\delta (\boldsymbol{p}),
\label{OmegaA}
\end{equation}%
and%
\begin{equation}
\int d^{2}p\,\langle m|e^{-i\boldsymbol{pX}}|n\rangle \langle i|e^{i\boldsymbol{pX}%
}|j\rangle =\frac{2\pi }{\ell _{B}^{2}}\delta _{ni}\delta _{mj}.
\label{OmegaOmega}
\end{equation}%
They are proved in Appendix A.

We derive the inversion relation of the Weyl ordering (\ref{WeylTrans}). We
evaluate%
\begin{align}
& \int \!d^{2}p\,\text{Tr}\left[ \widehat{O}_{f}e^{i\boldsymbol{p}(\boldsymbol{x}-%
\boldsymbol{X})}\right]  \notag \\
& =\frac{1}{(2\pi )^{2}}\int \!d^{2}pd^{2}qd^{2}y\,\text{Tr}\left[ e^{i%
\boldsymbol{qX}}e^{-i\boldsymbol{pX}}\right] e^{-\boldsymbol{qy}+i\boldsymbol{px}}f(\boldsymbol{y%
}).
\end{align}%
Using (\ref{MagneTrans}) and (\ref{OmegaA}), we obtain%
\begin{equation}
f(\boldsymbol{x})=\frac{\ell _{B}^{2}}{2\pi }\int \!d^{2}p\,e^{i\boldsymbol{px}}%
\text{Tr}\left[ \widehat{O}_{f}e^{-i\boldsymbol{pX}}\right] .  \label{WeylInver}
\end{equation}%
This is the inversion relation of the Weyl ordering. An interesting relation
follows trivially,%
\begin{equation}
\text{Tr}\left[ \widehat{O}_{f}\right] =\frac{1}{2\pi \ell _{B}^{2}}\int
\!d^{2}x\,f(\boldsymbol{x}).
\end{equation}%
We may regard this as a generalization of (\ref{OmegaA}), which is
reproduced by setting $f(\boldsymbol{x})=e^{i\boldsymbol{px}}$ and $\widehat{O}%
_{f}=e^{i\boldsymbol{pX}}$.

In this paper we deal with the second-quantized density operator. It is
necessary to define the Weyl ordering of such an operator. As a standard
procedure we proceed from \textit{classical mechanics} (CM) to \textit{%
quantum mechanics} (QM) and then to \textit{field theory} (FT)\cite%
{Eliashvili02}. The classical density is%
\begin{equation}
\rho _{\boldsymbol{r}}^{\text{CM}}(\boldsymbol{x})=\delta (\boldsymbol{x}-\boldsymbol{r}),
\end{equation}%
where $\boldsymbol{x}$ denotes the particle coordinate, $\boldsymbol{x}=\boldsymbol{x}%
(t) $: On the other hand, $\boldsymbol{r}$ is the variable parametrizing the
plane, which remains to be commutative after the first quantization as well
as the second quantization. Passing to quantum mechanics we replace the
c-number coordinate $\boldsymbol{x}$ by the corresponding operator $\boldsymbol{X}$.
Every quantity $f(\boldsymbol{x})$ is to be replaced by the Weyl-ordered one $%
\widehat{O}_{f}$ according to (\ref{WeylTrans}). Setting $f(\boldsymbol{x})=\rho
_{\boldsymbol{r}}^{\text{CM}}(\boldsymbol{x})=\delta (\boldsymbol{x}-\boldsymbol{r})$, we
find 
\begin{align}
\rho _{\boldsymbol{r}}^{\text{QM}}& \equiv \widehat{O}_{f}=\frac{1}{(2\pi )^{2}}%
\int \!d^{2}qd^{2}x\,e^{-i\boldsymbol{q}(\boldsymbol{x}-\boldsymbol{X})}\delta (\boldsymbol{x%
}-\boldsymbol{r})  \notag \\
& =\frac{1}{(2\pi )^{2}}\int \!d^{2}q\,e^{-i\boldsymbol{q}(\boldsymbol{r}-\boldsymbol{X}%
)}.
\end{align}%
Here, no requirement has yet been made on the commutativity of the operator $%
\boldsymbol{X}=(X,Y)$.

In passing to field theory, denoting by $|n\rangle $ the quantum mechanical
one-body state, we introduce the second-quantized fermion operator $c(n)$
with $\{c(n),c^{\dagger }(m)\}=\delta _{nm}$. We define%
\begin{equation}
|\Psi \rangle =\sum_{n}|n\rangle c(n),  \label{FieldExpanA}
\end{equation}%
so that the field operator is $\Psi (\boldsymbol{x})=\langle \boldsymbol{x}|\Psi
\rangle $. The field theoretical density operator is%
\begin{align}
\rho ^{\text{FT}}(\boldsymbol{r})& \equiv \langle \Psi |\rho _{\boldsymbol{r}}^{%
\text{QM}}|\Psi \rangle =\frac{1}{(2\pi )^{2}}\int \!d^{2}q\,e^{-i\boldsymbol{qr}%
}\langle \Psi |e^{i\boldsymbol{qX}}|\Psi \rangle  \notag \\
& =\frac{1}{(2\pi )^{2}}\int \!d^{2}qd^{2}xd^{2}y\,e^{-i\boldsymbol{qr}}\langle
\Psi |\boldsymbol{x}\rangle \langle \boldsymbol{x}|e^{i\boldsymbol{qX}}|\boldsymbol{y}%
\rangle \langle \boldsymbol{y}|\Psi \rangle .  \label{WeylTrans2nd}
\end{align}%
This is the standard procedure for second quantization, where the
first-quantized operator is sandwiched between $\Psi ^{\dag }(\boldsymbol{x})$
and $\Psi (\boldsymbol{x})$.

In order to show that the formula (\ref{WeylTrans2nd}) is the general one,
we first apply it to the commutative theory with $[X,Y]=0$. It is
represented by $\boldsymbol{X}|\boldsymbol{x}\rangle =\boldsymbol{x}|\boldsymbol{x}\rangle
\quad $with\quad $\langle \boldsymbol{x}|\boldsymbol{y}\rangle =\delta (\boldsymbol{x}-%
\boldsymbol{y})$. Using this in (\ref{WeylTrans2nd}) we find $\rho ^{\text{FT}}(%
\boldsymbol{r})=\rho (\boldsymbol{r})$ with%
\begin{equation}
\rho (\boldsymbol{r})=\Psi ^{\dag }(\boldsymbol{r})\Psi (\boldsymbol{r}).
\end{equation}%
It is the well-known result in the commutative theory.

We proceed to discuss the noncommutative theory with $[X,Y]=-i\ell _{B}^{2}$%
. It is represented by the Fock space made of (\ref{LandaSite}).
Substituting the expansion (\ref{FieldExpanA}) into (\ref{WeylTrans2nd}), we
obtain $\rho ^{\text{FT}}(\boldsymbol{r})=\hat{\rho}(\boldsymbol{r})$ with%
\begin{equation}
\hat{\rho}(\boldsymbol{r})=\frac{1}{(2\pi )^{2}}\int \!d^{2}q\,e^{-i\boldsymbol{qr}}%
\left[ \sum_{mn}\langle m|e^{i\boldsymbol{qX}}|n\rangle \rho (m,n)\right] ,
\label{WeylOrderDensiX}
\end{equation}%
where $\rho (m,n)\equiv c^{\dagger }(m)c(n)$. It approaches the ordinary
density $\rho (\boldsymbol{r})$ in the commutative limit ($\ell _{B}\rightarrow
0 $). The Fourier transformation is 
\begin{equation}
\hat{\rho}(\boldsymbol{q})=\frac{1}{2\pi }\sum_{mn}\langle m|e^{-i\boldsymbol{qX}%
}|n\rangle \rho (m,n).  \label{WeylOrderDensiQ}
\end{equation}%
We propose $\hat{\rho}(\boldsymbol{r})$ as the Weyl-ordered density operator. We
interpret that the matrix element $\langle \hat{\rho}(\boldsymbol{r})\rangle $
is the classical density measured at the point $\boldsymbol{r}$\ in the
commutative space. The commutative space is the one from which the
noncommutative space is constructed by deforming the commutation relation:
It is used for the representation (\ref{RepreDiffe}) of the noncommutativity
(\ref{NonCommuXY}).

\section{Lowest Landau Level Projection}

We proceed to discuss QH systems. Electrons make cyclotron motions under
perpendicular magnetic field $B$ and their energies are quantized into
Landau levels. The number density of magnetic flux quanta is $\rho _{\Phi
}\equiv B/\Phi _{\text{D}}$, with $\Phi _{\text{D}}=2\pi \hbar /e$ the flux
unit, which is equal to the number density of Landau sites. One electron
occupies area $2\pi \ell _{B}^{2}$ with $\ell _{B}=\sqrt{\hbar /eB}$ the
magnetic length. The filling factor is $\nu =\rho _{0}/\rho _{\Phi }$ with $%
\rho _{0}$ the electron number density. At $\nu =k$ (integer), one Landau
site accommodates $k$ electrons with different isospins due to the Pauli
exclusion principle.

We first review the one-body property of electrons in strong magnetic field.
The electron coordinate $\boldsymbol{x}=(x,y)$ is decomposed as $\boldsymbol{x}=%
\boldsymbol{X}+\boldsymbol{R}$, where $\boldsymbol{X}=(X,Y)$ is the guiding center and $%
\boldsymbol{R}=(-P_{y},P_{x})/eB$ is the relative coordinate. From them we
construct two sets of harmonic-oscillator operators, 
\begin{subequations}
\begin{align}
& a\equiv {\frac{\ell _{B}}{\sqrt{2}\hbar }}(P_{x}+iP_{y}),\quad a^{\dagger
}\equiv {\frac{\ell _{B}}{\sqrt{2}\hbar }}(P_{x}-iP_{y}),  \label{OperaA} \\
& b\equiv {\frac{1}{\sqrt{2}\ell _{B}}}(X-iY),\quad b^{\dagger }\equiv {%
\frac{1}{\sqrt{2}\ell _{B}}}(X+iY),  \label{OperaB}
\end{align}%
obeying 
\end{subequations}
\begin{equation}
\lbrack a,a^{\dagger }]=[b,b^{\dagger }]=1,\qquad \lbrack a,b]=[a^{\dagger
},b]=0.  \label{HarmoCommuAB}
\end{equation}%
The kinetic Hamiltonian is 
\begin{equation}
H_{\text{K}}=(a^{\dagger }a+{\frac{1}{2}})\hbar \omega _{\text{c}}
\label{HamilKinemQM}
\end{equation}%
with $\hbar \omega _{\text{c}}$ the cyclotron energy. When the cyclotron gap
is large enough, thermal excitations across Landau levels are practically
impossible. Hence, it is a good approximation to neglect all those
excitations by requiring the confinement of electrons to the LLL. The
guiding center is the noncommutative coordinate.

We make the LLL projection in a systematic way\cite{BookEzawa}. We decompose
the coordinate $\boldsymbol{x}$ into the relative coordinate $\boldsymbol{R}$ and
the guiding center $\boldsymbol{X}$. The relative coordinate $\boldsymbol{R}$ is
frozen when the electron is confined to the LLL. We denote the LLL
projection of the c-number function $f(\boldsymbol{x})$ as $\langle \kern%
-1mm\langle f\rangle \kern-1mm\rangle $. In particular, we have\cite%
{BookEzawa}%
\begin{equation}
\langle \kern-1mm\langle f\rangle \kern-1mm\rangle =e^{-\ell _{B}^{2}\boldsymbol{%
p}^{2}/4}e^{i\boldsymbol{pX}}  \label{ProjeWaveX}
\end{equation}%
for the plane wave $f(x)=e^{i\boldsymbol{px}}$. The suppression factor $e^{-\ell
_{B}^{2}\boldsymbol{p}^{2}/4}$ arises due to the LLL projection of the relative
coordinate. In general, we have%
\begin{equation}
\langle \kern-1mm\langle f\rangle \kern-1mm\rangle =\frac{1}{(2\pi )^{2}}%
\int \!d^{2}qd^{2}x\,e^{-\ell _{B}^{2}\boldsymbol{q}^{2}/4}e^{-i\boldsymbol{q}(%
\boldsymbol{x}-\boldsymbol{X})}f(\boldsymbol{x}).  \label{ProjeLLLf}
\end{equation}%
Consequently, the LLL projection is equivalent to the Weyl ordering (\ref%
{WeylTrans}) but for the suppression factor.

In the field theoretical framework the kinetic Hamiltonian (\ref%
{HamilKinemQM}) reads 
\begin{equation}
H_{\text{K}}={\frac{1}{2M}}\int d^{2}x\boldsymbol{\psi }^{\dagger }(\boldsymbol{x}%
)(P_{x}-iP_{y})(P_{x}+iP_{y})\boldsymbol{\psi }(\boldsymbol{x}),  \label{HamilKinem}
\end{equation}%
apart from the cyclotron energy $\hbar \omega _{\text{c}}/2$ per electron,
where $P_{k}=-i\hbar \partial _{k}+eA_{k}$. We assume the electron field $%
\boldsymbol{\psi }$ to possess $N$ isospin components $\psi _{\sigma }$. We
introduce the field operator describing electrons confined to the LLL. It is
determined so as to satisfy the LLL condition%
\begin{equation}
(P_{x}+iP_{y})\psi _{\sigma }(\boldsymbol{x})|\mathfrak{S}\rangle =0,
\label{LLLcondi}
\end{equation}%
implying that the kinetic Hamiltonian (\ref{HamilKinem}) is quenched.
Solving this equation we find the projected field to be 
\begin{equation}
\bar{\psi}_{\sigma }(\boldsymbol{x})=\sum_{n=0}^{\infty }c_{\sigma }(n)\langle 
\boldsymbol{x}|n\rangle ,  \label{FieldExpan}
\end{equation}%
where $\langle \boldsymbol{x}|n\rangle $ is the one-body wave function (\ref%
{OneBodyLLL}) and $c_{\sigma }(n)$ is the annihilation operator of the
electron with isospin $\sigma $ at the Landau site $n$, 
\begin{align}
\{c_{\sigma }(n),c_{\tau }^{\dagger }(m)\}& =\delta _{nm}\delta _{\sigma
\tau },  \notag \\
\{c_{\sigma }(n),c_{\tau }(m)\}& =\{c_{\sigma }^{\dagger }(n),c_{\tau
}^{\dagger }(m)\}=0.  \label{AntiCommuC}
\end{align}%
The Hilbert space $\mathbb{H}_{\text{LLL}}$ is made of the Fock spaces $%
\mathbb{H}_{n}$ of electrons in all Landau sites, $\mathbb{H}_{\text{LLL}%
}=\otimes _{n}\mathbb{H}_{n}$.

The LLL projection of the density operator $\rho (\boldsymbol{x})=\boldsymbol{\psi }%
^{\dagger }(\boldsymbol{x})\boldsymbol{\psi }(\boldsymbol{x})$ is given by\cite%
{Iso92PLB,Cappelli93NPB}%
\begin{equation}
\bar{\rho}(\boldsymbol{x})=\boldsymbol{\bar{\psi}}^{\dagger }(\boldsymbol{x})\boldsymbol{%
\bar{\psi}}(\boldsymbol{x}).  \label{ProjeDensiNaive}
\end{equation}%
Substituting the expansion (\ref{FieldExpan}) into it we find 
\begin{equation}
\bar{\rho}(\boldsymbol{x})=\sum_{mn}\langle m|\boldsymbol{x\rangle }\langle \boldsymbol{x%
}|n\rangle \rho (m,n),  \label{RhoStepA}
\end{equation}%
where 
\begin{equation}
\rho (m,n)\equiv \sum_{\sigma }c_{\sigma }^{\dagger }(m)c_{\sigma }(n).
\end{equation}%
The Fourier transformation of $\bar{\rho}(\boldsymbol{x})$ is%
\begin{equation}
\bar{\rho}(\boldsymbol{q})=\sum_{mn}^{\infty }\rho (m,n)\int \frac{d^{2}x}{2\pi }%
\,e^{-i\boldsymbol{qx}}\langle m|\boldsymbol{x}\rangle \langle \boldsymbol{x}|n\rangle .
\label{FouriLLL}
\end{equation}%
Since $e^{-i\boldsymbol{px}}$ is just a c-number, this is moved into the matrix
element. We replace $\boldsymbol{x}$ with the position operator acting on the
state $|\boldsymbol{x}\rangle $, and separate it into the guiding center $%
\boldsymbol{X}$\ and the relative coordinate $\boldsymbol{R}$. The relative
coordinate being frozen, the plane wave $e^{-i\boldsymbol{qx}}$ is projected as
in (\ref{ProjeWaveX}), 
\begin{equation}
\int \!d^{2}x\,e^{-i\boldsymbol{qx}}\langle m|\boldsymbol{x}\rangle \langle \boldsymbol{x%
}|n\rangle =e^{-\ell _{B}^{2}\boldsymbol{q}^{2}/4}\langle m|e^{-i\boldsymbol{qX}%
}|n\rangle ,  \label{ProjeStepX}
\end{equation}%
where we have used $\int d^{2}x\,|\boldsymbol{x}\rangle \langle \boldsymbol{x}|=1$.
Hence, we find%
\begin{equation}
\bar{\rho}(\boldsymbol{q})=\frac{1}{2\pi }e^{-\ell _{B}^{2}\boldsymbol{q}%
^{2}/4}\sum_{mn}^{\infty }\langle m|e^{-i\boldsymbol{qX}}|n\rangle \rho (m,n).
\label{ProjeDensiPre}
\end{equation}%
It is equivalent to the Weyl-ordered density operator $\hat{\rho}(\boldsymbol{q}%
) $ given by (\ref{WeylOrderDensiQ}) but for the suppression factor. This is
what we have expected from (\ref{WeylTrans}) and (\ref{ProjeLLLf}).

Let us reexamine the LLL projection of the density operator in a spirit of
the basic formula (\ref{ProjeWaveX}). We start with the Fourier
transformation of the unprojected density, 
\begin{subequations}
\begin{align}
\rho (\boldsymbol{q})& =\int \frac{d^{2}x}{2\pi }\,e^{-i\boldsymbol{qx}}\psi
_{\sigma }^{\dagger }(\boldsymbol{x})\psi _{\sigma }(\boldsymbol{x})
\label{DensiStepAx} \\
& =\int \frac{d^{2}x}{2\pi }\int d^{2}y\,\psi _{\sigma }^{\dagger }(\boldsymbol{x%
})\langle \boldsymbol{x}|e^{-i\boldsymbol{qx}}|\boldsymbol{y}\rangle \psi _{\sigma }(%
\boldsymbol{y}).  \label{DensiStepA}
\end{align}%
Here, we project the plane wave according to (\ref{ProjeWaveX}), 
\end{subequations}
\begin{equation}
\langle \kern-1mm\langle \rho (\boldsymbol{q})\rangle \kern-1mm\rangle =e^{-\ell
_{B}^{2}\boldsymbol{q}^{2}/4}\int \frac{d^{2}xd^{2}y}{2\pi }\,\langle \boldsymbol{x}%
|e^{-i\boldsymbol{qX}}|\boldsymbol{y}\rangle \psi _{\sigma }^{\dagger }(\boldsymbol{x}%
)\psi _{\sigma }(\boldsymbol{y}),  \label{DensiStepB}
\end{equation}%
as is done by substituting the completeness condition in the Hilbert space $%
\mathbb{H}_{\text{LLL}}$, $\sum_{m=0}^{\infty }|m\rangle \langle m|=1$.
Then, it is easy to see that the formula (\ref{DensiStepB}) is reduced to (%
\ref{ProjeDensiPre}).

A comment is in order. When we take the LLL projection of the plane wave
naively in (\ref{DensiStepAx}), we would obtain 
\begin{equation}
\langle \kern-1mm\langle \rho (\boldsymbol{q})\rangle \kern-1mm\rangle =e^{-\ell
_{B}^{2}\boldsymbol{q}^{2}/4}\int \frac{d^{2}x}{2\pi }\,e^{-i\boldsymbol{qX}}\rho (%
\boldsymbol{x}).
\end{equation}%
This is the formula given in Ref.\cite{BookEzawa}. It should be understood
as a symbolic notation of (\ref{DensiStepB}).

\section{Noncommutative Algebra}

The kinetic Hamiltonian (\ref{HamilKinem}) possesses the global symmetry
U(N)=U(1)$\otimes $SU(N), whose generators are%
\begin{equation}
\rho (\boldsymbol{x})=\boldsymbol{\psi }^{\dagger }(\boldsymbol{x})\boldsymbol{\psi }(%
\boldsymbol{x}),\quad S^{A}(\boldsymbol{x})=\frac{1}{2}\boldsymbol{\psi }^{\dagger }(%
\boldsymbol{x})\lambda ^{A}\boldsymbol{\psi }(\boldsymbol{x}),  \label{GenerUN}
\end{equation}%
where $\lambda ^{A}$ are the generating matrices, 
\begin{align}
\lbrack \lambda ^{A},\lambda ^{B}]& =2if^{ABC}\lambda ^{C},  \notag \\
\{\lambda ^{A},\lambda ^{B}\}& =2d^{ABC}\lambda ^{C}+\frac{4}{N}\delta ^{AB},
\label{LambdCommu}
\end{align}%
with $f^{ABC}$ and $d^{ABC}$ the structure constants of SU(N).

We investigate how the symmetry is modified for those electrons in the
noncommutative plane. In the momentum space the Weyl-ordered generators are
given by (\ref{WeylOrderDensiQ}), or%
\begin{align}
\hat{\rho}(\boldsymbol{q})& =\frac{1}{2\pi }\sum_{mn}^{\infty }\langle m|e^{-i%
\boldsymbol{qX}}|n\rangle \rho (m,n),  \label{ProjeDensiP} \\
\hat{S}^{A}(\boldsymbol{q})& =\frac{1}{2\pi }\sum_{mn}^{\infty }\langle m|e^{-i%
\boldsymbol{qX}}|n\rangle S^{A}(m,n),  \label{ProjeDensiS}
\end{align}%
with%
\begin{align}
\rho (m,n)& \equiv \sum_{\sigma }c_{\sigma }^{\dagger }(m)c_{\sigma }(n),
\label{Dij} \\
S^{A}(m,n)& \equiv \frac{1}{2}\sum_{\sigma \tau }c_{\sigma }^{\dagger
}(m)\lambda _{\sigma \tau }^{A}c_{\tau }(n).  \label{Sij}
\end{align}%
Taking the Fourier transformation we have%
\begin{align}
\hat{\rho}(\boldsymbol{x})=& \int \frac{d^{2}q}{2\pi }\,e^{i\boldsymbol{qx}}\hat{\rho%
}(\boldsymbol{q}),  \label{ProjeDensiX} \\
\hat{S}^{A}(\boldsymbol{x})=& \int \frac{d^{2}q}{2\pi }\,e^{i\boldsymbol{qx}}\hat{S}%
^{A}(\boldsymbol{q}).
\end{align}%
The inversions of (\ref{ProjeDensiP}) and (\ref{ProjeDensiS}) are 
\begin{align}
\rho (m,n)& =\ell _{B}^{2}\int d^{2}q\,\langle n|e^{i\boldsymbol{qX}}|m\rangle 
\hat{\rho}(\boldsymbol{q}),  \label{DijQ} \\
S^{A}(m,n)& =\ell _{B}^{2}\int d^{2}q\,\langle n|e^{i\boldsymbol{qX}}|m\rangle 
\hat{S}^{A}(\boldsymbol{q}),  \label{SijQ}
\end{align}%
as is verified with the use of (\ref{OmegaOmega}).

The operators $\rho (m,n)$ generate the algebra 
\begin{equation}
\lbrack \rho (m,n),\rho (j,k)]=\delta _{jn}\rho (m,k)-\delta _{mk}\rho (j,n),
\label{WinftyR}
\end{equation}%
as follows from the anticommutation relation (\ref{AntiCommuC}) of $%
c_{\sigma }(m)$. This is closely related to the W$_{\infty }$ algebra (\ref%
{WinftyB}). It is easy to show that the element 
\begin{equation}
\mathcal{L}(m,n)=\int d^{2}x\,\boldsymbol{\bar{\psi}}^{\dagger }(\boldsymbol{x}%
)L(m,n)\boldsymbol{\bar{\psi}}(\boldsymbol{x})
\end{equation}%
generates the algebra isomorphic to (\ref{WinftyB}). On the other hand, they
span the same linear space as $\rho (m,n)$ span: $\mathcal{L}%
(0,0)=\sum_{m}\rho (m,m)$, $\mathcal{L}(0,1)=\sum_{m}\sqrt{m+1}\rho (m,m+1)$%
, $\mathcal{L}(1,1)=\sum_{m}m\rho (m,m)$, and so on. Hence $\mathcal{L}(m,n)$
and $\rho (m,n)$ give the same Fock representation of W$_{\infty }$.

The set of $\rho (m,n)$ and $S^{A}(m,n)$ generate an extended algebra. 
\begin{align}
& [\rho (m,n),\rho (j,k)]=\delta _{jn}\rho (m,k)-\delta _{mk}\rho (j,n), 
\notag \\
& [\rho (m,n),S^{A}(j,k)]=\delta _{jn}S^{A}(m,k)-\delta _{mk}S^{A}(j,n), 
\notag \\
& [S^{A}(m,n),S^{B}(j,k)]  \notag \\
& \hspace{0.7cm}=\frac{i}{2}f^{ABC}\left[ \delta _{jn}S^{C}(m,k)+\delta
_{mk}S^{C}(j,n)\right]  \notag \\
& \hspace{0.7cm}+\frac{1}{2}d^{ABC}\left[ \delta _{jn}S^{C}(m,k)-\delta
_{mk}S^{C}(j,n)\right]  \notag \\
& \hspace{0.7cm}+\frac{1}{2N}\delta ^{AB}\left[ \delta _{jn}\rho
(m,k)-\delta _{mk}\rho (j,n)\right] ,  \label{WalgebP}
\end{align}%
as follows from the anticommutation relations (\ref{AntiCommuC}) of $%
c_{\sigma }(m)$. We reformulate it in terms of the electron density $\hat{%
\rho}(\boldsymbol{p})$ and the isospin density $\hat{S}^{A}(\boldsymbol{p})$, 
\begin{align}
& [\hat{\rho}(\boldsymbol{p}),\hat{\rho}(\boldsymbol{q})]={\frac{i}{\pi }}\hat{\rho}(%
\boldsymbol{p}+\boldsymbol{q})\sin \left[ \frac{1}{2}\ell _{B}^{2}\boldsymbol{p}\!\wedge
\!\boldsymbol{q}\right] ,  \notag \\
& [\hat{S}^{A}(\boldsymbol{p}),\hat{\rho}(\boldsymbol{q})]={\frac{i}{\pi }}\hat{S}%
^{A}(\boldsymbol{p}+\boldsymbol{q})\sin \left[ \frac{1}{2}\ell _{B}^{2}\boldsymbol{p}%
\!\wedge \!\boldsymbol{q}\right] ,  \notag \\
& [\hat{S}^{A}(\boldsymbol{p}),\hat{S}^{B}(\boldsymbol{q})]={\frac{i}{2\pi }}f^{ABC}%
\hat{S}^{C}(\boldsymbol{p}+\boldsymbol{q})\cos \left[ \frac{1}{2}\ell _{B}^{2}%
\boldsymbol{p}\!\wedge \!\boldsymbol{q}\right]  \notag \\
& \hspace{0.7in}+{\frac{i}{2\pi }}d^{ABC}\hat{S}^{C}(\boldsymbol{p}+\boldsymbol{q}%
)\sin \left[ \frac{1}{2}\ell _{B}^{2}\boldsymbol{p}\!\wedge \!\boldsymbol{q}\right] 
\notag \\
& \hspace{0.7in}+{\frac{i}{2\pi N}}\delta ^{AB}\hat{\rho}(\boldsymbol{p}+\boldsymbol{%
q})\sin \left[ \frac{1}{2}\ell _{B}^{2}\boldsymbol{p}\!\wedge \!\boldsymbol{q}\right]
.  \label{WAlgeb}
\end{align}%
These are easily derived with the use of (\ref{ProjeDensiP}), (\ref%
{ProjeDensiS}), (\ref{WalgebP}) and (\ref{MagneTrans}). We call (\ref%
{WalgebP}) or equivalently (\ref{WAlgeb}) the W$_{\infty }$(N) algebra since
it is the SU(N) extension of W$_{\infty }$.

In the coordinate space the commutation relations read 
\begin{align}
& [\hat{\rho}(\boldsymbol{x}),\hat{\rho}(\boldsymbol{y})]=\int d^{2}z\,[\kern%
-0.5mm[\delta _{\boldsymbol{x}}(\boldsymbol{z}),\delta _{\boldsymbol{y}}(\boldsymbol{z})]%
\kern-0.5mm]\,\hat{\rho}(\boldsymbol{z}),  \notag \\
& [\hat{S}^{A}(\boldsymbol{x}),\hat{\rho}(\boldsymbol{y})]=\int d^{2}z\,[\kern%
-0.5mm[\delta _{\boldsymbol{x}}(\boldsymbol{z}),\delta _{\boldsymbol{y}}(\boldsymbol{z})]%
\kern-0.5mm]\,\hat{S}^{A}(\boldsymbol{z}),  \notag \\
& [\hat{S}^{A}(\boldsymbol{x}),\hat{S}^{B}(\boldsymbol{y})]={\frac{i}{2}}f^{ABC}\int
d^{2}z\,\{\kern-1mm\{\delta _{\boldsymbol{x}}(\boldsymbol{z}),\delta _{\boldsymbol{y}}(%
\boldsymbol{z})\}\kern-1mm\}\,\hat{S}^{C}(\boldsymbol{z})  \notag \\
& \hspace*{0.3in}+{\frac{1}{2}}d^{ABC}\int d^{2}z\,[\kern-0.5mm[\delta _{%
\boldsymbol{x}}(\boldsymbol{z}),\delta _{\boldsymbol{y}}(\boldsymbol{z})]\kern-0.5mm]\,\hat{S%
}^{C}(\boldsymbol{z})  \notag \\
& \hspace*{0.3in}+{\frac{1}{2N}}\delta ^{AB}\int d^{2}z\,[\kern-0.5mm[\delta
_{\boldsymbol{x}}(\boldsymbol{z}),\delta _{\boldsymbol{y}}(\boldsymbol{z})]\kern-0.5mm]\,%
\hat{\rho}(\boldsymbol{z}),
\end{align}%
where $[\kern-0.5mm[\delta _{\boldsymbol{x}}(\boldsymbol{z}),\delta _{\boldsymbol{y}}(%
\boldsymbol{z})]\kern-0.5mm]$ is the Moyal bracket, 
\begin{align}
\lbrack \kern-0.5mm[& \delta _{\boldsymbol{x}}(\boldsymbol{z}),\delta _{\boldsymbol{y}}(%
\boldsymbol{z})]\kern-0.5mm]  \notag \\
& =\delta _{\boldsymbol{x}}(\boldsymbol{z})\bigstar \delta _{\boldsymbol{y}}(\boldsymbol{z}%
)-\delta _{\boldsymbol{y}}(\boldsymbol{z})\bigstar \delta _{\boldsymbol{x}}(\boldsymbol{z}) 
\notag \\
& =\frac{2i}{(2\pi )^{4}}\int d^{2}pd^{2}q\,e^{i\boldsymbol{p}(\boldsymbol{x}-%
\boldsymbol{z})+i\boldsymbol{q}(\boldsymbol{y}-\boldsymbol{z})}\sin \left[ \frac{1}{2}\ell
_{B}^{2}\boldsymbol{p}\!\wedge \!\boldsymbol{q}\right] ,
\end{align}%
and $\{\kern-1mm\{\delta _{\boldsymbol{x}}(\boldsymbol{z}),\delta _{\boldsymbol{y}}(%
\boldsymbol{z})\}\kern-1mm\}$ is the Moyal antibracket, 
\begin{align}
\{\kern-1mm\{& \delta _{\boldsymbol{x}}(\boldsymbol{z}),\delta _{\boldsymbol{y}}(\boldsymbol{%
z})\}\kern-1mm\}  \notag \\
& =\delta _{\boldsymbol{x}}(\boldsymbol{z})\bigstar \delta _{\boldsymbol{y}}(\boldsymbol{z}%
)+\delta _{\boldsymbol{y}}(\boldsymbol{z})\bigstar \delta _{\boldsymbol{x}}(\boldsymbol{z}) 
\notag \\
& =\frac{2}{(2\pi )^{4}}\int d^{2}pd^{2}q\,e^{i\boldsymbol{p}(\boldsymbol{x}-\boldsymbol{%
z})+i\boldsymbol{q}(\boldsymbol{y}-\boldsymbol{z})}\cos \left[ \frac{1}{2}\ell _{B}^{2}%
\boldsymbol{p}\!\wedge \!\boldsymbol{q}\right] .
\end{align}%
Here, $\delta _{\boldsymbol{x}}(\boldsymbol{z})=\delta (\boldsymbol{x}-\boldsymbol{z})$, and 
$\bigstar $ denotes the star-product with respect to $\boldsymbol{z}$. We have
adopted the convention%
\begin{equation}
f(\boldsymbol{z})\bigstar g(\boldsymbol{z})=\exp \left( -\frac{i}{2}\ell
_{B}^{2}\bigtriangledown _{\boldsymbol{x}}\wedge \bigtriangledown _{\boldsymbol{y}%
}\right) f(\boldsymbol{x})g(\boldsymbol{y}){\Large |}_{\boldsymbol{x}=\boldsymbol{y}=\boldsymbol{%
z}}
\end{equation}%
in accord with the noncommutativity (\ref{NonCommuXY}).

The algebra W$_{\infty }$(N) regresses to the algebra U(N) in the
commutative limit ($\ell _{B}\rightarrow 0$), where%
\begin{align}
\lbrack \kern-0.5mm[\delta _{\boldsymbol{x}}(\boldsymbol{z}),\delta _{\boldsymbol{y}}(%
\boldsymbol{z})]\kern-0.5mm]& \rightarrow 0,  \notag \\
\{\kern-1mm\{\delta _{\boldsymbol{x}}(\boldsymbol{z}),\delta _{\boldsymbol{y}}(\boldsymbol{z}%
)\}\kern-1mm\}& \rightarrow 2\delta (\boldsymbol{x}-\boldsymbol{z})\delta (\boldsymbol{y}%
-\boldsymbol{z}).  \label{CommuLimit}
\end{align}%
In this limit the densities $\hat{\rho}(\boldsymbol{x})$ and $\hat{S}^{A}(%
\boldsymbol{x})$ are reduced to the physical densities $\rho (\boldsymbol{x})$ and $%
S^{A}(\boldsymbol{x})$\ in the original commutative geometry.

It is instructive to evaluate the commutation relations of the projected
densities (\ref{ProjeDensiNaive}). From (\ref{WAlgeb}) we obtain%
\begin{equation}
\lbrack \bar{\rho}(\boldsymbol{x}),\bar{\rho}(\boldsymbol{y})]=\int d^{2}z\,K^{-}(%
\boldsymbol{x,y;z})\bar{\rho}(\boldsymbol{z}),  \label{RhoCommuUn}
\end{equation}%
and so on, where the kernel $K^{-}(\boldsymbol{x,y;z})$ contains the integration%
\begin{equation}
\int d^{2}pd^{2}q\,e^{i\boldsymbol{p}(\boldsymbol{x}-\boldsymbol{z})+i\boldsymbol{q}(\boldsymbol{%
y}-\boldsymbol{z})+(i\ell _{B}^{2}/2)\boldsymbol{p}\wedge \boldsymbol{q}+(\ell
_{B}^{2}/2)\boldsymbol{pq}}.
\end{equation}%
It is divergent due to the factor $e^{(\ell _{B}^{2}/2)\boldsymbol{pq}}$.
Similar divergences appear also in $[\bar{S}^{A}(\boldsymbol{x}),\bar{\rho}(%
\boldsymbol{y})]$ and $[\bar{S}^{A}(\boldsymbol{x}),\bar{S}^{B}(\boldsymbol{y})]$. The
projected densities $\bar{\rho}(\boldsymbol{x})$ and $\bar{S}^{A}(\boldsymbol{x})$
are not good operators with respect to the W$_{\infty }$(N)\ algebra.

\section{Coulomb Interactions}

Electrons interact with each other via the Coulomb potential, 
\begin{equation}
H_{\text{C}}={\frac{1}{2}}\int d^{2}xd^{2}y\,\rho (\boldsymbol{x})V(\boldsymbol{x}-%
\boldsymbol{y})\rho (\boldsymbol{y}),  \label{CouloHamilOri}
\end{equation}%
where $V(\boldsymbol{x}-\boldsymbol{y})={e^{2}}/4\pi \varepsilon |\boldsymbol{x}-\boldsymbol{%
y}|$. (We later include the Zeeman and tunneling interactions to discuss
realistic systems.) In the previous sections we have projected the states to
the LLL. However, even if we start with a state in the LLL, the potential
term kicks it out up to higher Landau levels and results in an increase of
the kinetic energy. To suppress such excitations we make the LLL projection%
\cite{Girvin84B,Girvin86B} of the potential term additionally.

The projected Coulomb Hamiltonian is given by replacing the density $\rho (%
\boldsymbol{x})$ with the projected density $\bar{\rho}(\boldsymbol{x})$,

\begin{equation}
\hat{H}_{\text{C}}={\frac{1}{2}}\int d^{2}xd^{2}y\,\bar{\rho}(\boldsymbol{x})V(%
\boldsymbol{x}-\boldsymbol{y})\bar{\rho}(\boldsymbol{y}).  \label{CouloHamilLLL}
\end{equation}%
We substitute the expansion (\ref{FieldExpan}) of the electron field into
the projected density and reduce (\ref{CouloHamilLLL}) into 
\begin{equation}
\hat{H}_{\text{C}}=\sum_{mnij}\sum_{\sigma \tau }{V}_{mnij}c_{\sigma
}^{\dagger }(m)c_{\tau }^{\dagger }(i)c_{\tau }(j)c_{\sigma }(n),
\label{HamilInC}
\end{equation}%
where 
\begin{equation}
{V}_{mnij}={\frac{1}{2}}\int d^{2}xd^{2}yV(\boldsymbol{x}-\boldsymbol{y})\langle m|%
\boldsymbol{x}\rangle \langle \boldsymbol{x}|n\rangle \langle i|\boldsymbol{y}\rangle
\langle \boldsymbol{y}|j\rangle .  \label{CouloIJKL}
\end{equation}%
By using (\ref{ProjeStepX}) this is rewritten as%
\begin{equation}
{V}_{mnij}={\frac{1}{4\pi }}\int d^{2}k\,e^{-\ell _{B}^{2}\boldsymbol{k}^{2}/2}V(%
\boldsymbol{k})\langle m|e^{i\boldsymbol{Xk}}|n\rangle \langle i|e^{-i\boldsymbol{Xk}%
}|j\rangle .  \label{Vmnij}
\end{equation}%
We may rewrite (\ref{HamilInC}) as%
\begin{equation}
\hat{H}_{\text{C}}=\sum_{mnij}{V}_{mnij}\rho (m,n)\rho (i,j)
\label{EquivHamilD}
\end{equation}%
with (\ref{Dij}).

In terms of the Weyl-ordered density, (\ref{CouloHamilLLL}) yields 
\begin{equation}
\hat{H}_{\text{C}}={\frac{1}{2}}\int d^{2}xd^{2}y\,V_{\text{D}}(\boldsymbol{x}-%
\boldsymbol{y})\hat{\rho}(\boldsymbol{x})\hat{\rho}(\boldsymbol{y}).  \label{DirecHamilX}
\end{equation}%
It is derived from the expression in the momentum space,%
\begin{equation}
\hat{H}_{\text{C}}=\pi \int d^{2}k\,V_{\text{D}}(\boldsymbol{k})\hat{\rho}(-%
\boldsymbol{k})\hat{\rho}(\boldsymbol{k}).  \label{DirecHamilMom}
\end{equation}%
Here, we have separated out the suppression factors from the density
operator $\bar{\rho}(\boldsymbol{k})$ in (\ref{ProjeDensiPre})\ and have
attached it to $V(\boldsymbol{k})$ to construct $V_{\text{D}}(\boldsymbol{k})$. The
potential is given by%
\begin{equation}
V_{\text{D}}(\boldsymbol{k})=e^{-\ell _{B}^{2}\boldsymbol{k}^{2}/2}V(\boldsymbol{k})
\label{PontiVD}
\end{equation}%
with $V(\boldsymbol{k})=e^{2}/4\pi \varepsilon |\boldsymbol{k}|$. Its Fourier
transformation is%
\begin{equation}
V_{\text{D}}(\boldsymbol{x})=\frac{e^{2}\sqrt{2\pi }}{8\pi \varepsilon \ell _{B}}%
I_{0}(\boldsymbol{x}^{2}/4\ell _{B}^{2})e^{-\boldsymbol{x}^{2}/4\ell _{B}^{2}},
\end{equation}
where $I_{0}(x)$ is the modified Bessel function. It approaches the ordinary
Coulomb potential at large distance,%
\begin{equation}
V_{\text{D}}(\boldsymbol{x})\rightarrow V(\boldsymbol{x})=\frac{e^{2}}{4\pi
\varepsilon |\boldsymbol{x}|}\quad \text{as}\quad |\boldsymbol{x}|\rightarrow \infty
,
\end{equation}%
but at short distance it does not diverge in contrast to the ordinary
Coulomb potential,%
\begin{equation}
V_{\text{D}}(\boldsymbol{x})\rightarrow \frac{e^{2}\sqrt{2\pi }}{8\pi
\varepsilon \ell _{B}}\quad \text{as}\quad |\boldsymbol{x}|\rightarrow 0.
\label{RegulCoulo}
\end{equation}%
This is physically reasonable because a real electron cannot be localized to
a point within the LLL. The regularity (\ref{RegulCoulo}) of the potential
is attributed to the exponential suppression factor in (\ref{PontiVD}),
whose origin is the suppression factor in the LLL projection (\ref%
{ProjeWaveX}).

We consider a local SU(N) isospin rotation of electrons. Without the LLL
projection, since the isospin generator commutes with the density operator,
it does not affect the Coulomb energy (\ref{CouloHamilOri}) but increases a
kinetic energy. However, since the kinetic energy is very large in high
magnetic field, it is energetically favorable for electrons to stay within
the LLL. Namely, the confinement of electrons to the LLL occurs dynamically.
This dynamical effect is taken into account automatically by making the LLL
projection.

With the LLL condition (\ref{LLLcondi}), the kinetic Hamiltonian (\ref%
{HamilKinem}) is quenched, $H_{\text{K}}|\mathfrak{S}\rangle =0$ for $|%
\mathfrak{S}\rangle \in \mathbb{H}_{\text{LLL}}$. Thus, a local SU(N)
isospin rotation takes place without requiring kinetic energy within the
LLL. However, it turns out to increase the Coulomb energy (\ref{DirecHamilX}%
) because an isospin rotation modulates the electron density according to
the algebra (\ref{WAlgeb}). It has been argued\cite%
{Moon95B,Ezawa97B,EzawaX02B} that this results in the increase of the
exchange Coulomb energy and leads to a new physics associated with quantum
coherence. However, the Hamiltonian has not yet been obtained in a closed
form, and it would be impossible in the previous methods. In one method\cite%
{Moon95B,Ezawa97B}, the effective Hamiltonian is extracted by evaluating the
Coulomb energy of a sufficiently smooth spin texture. In another method\cite%
{EzawaX02B}, it is constructed by taking a continuum limit of the
Landau-site Hamiltonian. In these methods it is difficult to calculate
higher order corrections systematically. In this section we propose an
algebraic analysis to overcome this problem.

The key observation is that the projected Coulomb Hamiltonian (\ref%
{CouloHamilLLL}) can be represented in an entirely different form. Making
use of the relation%
\begin{equation}
\sum_{A}^{N^{2}-1}\lambda _{\sigma \tau }^{A}\lambda _{\alpha \beta
}^{A}=2\left( \delta _{\sigma \beta }\delta _{\tau \alpha }-\frac{1}{N}%
\delta _{\sigma \tau }\delta _{\alpha \beta }\right) ,
\end{equation}%
we rewrite (\ref{HamilInC}) as

\begin{align}
\hat{H}_{\text{C}}& =-2\sum_{mnij}{V}_{mnij}[S^{A}(m,j)S^{A}(i,n)  \notag \\
& \hspace{0.5in}+\frac{1}{2N}\rho (m,j)\rho (i,n)].  \label{EquivHamilX}
\end{align}%
We substitute (\ref{DijQ}), (\ref{SijQ}) and (\ref{Vmnij}) into it, and use
the relation%
\begin{align}
& \sum_{n}\langle n|e^{-i\boldsymbol{Xk}}e^{i\boldsymbol{Xp}}e^{i\boldsymbol{Xk}}e^{i%
\boldsymbol{Xq}}|n\rangle   \notag \\
& \hspace{0.4in}=\frac{2\pi }{\ell _{B}^{2}}\delta (\boldsymbol{p}+\boldsymbol{q}%
)\exp \left( i\ell _{B}^{2}\boldsymbol{p}\wedge \boldsymbol{k}\right) ,
\end{align}%
which follows from (\ref{MagneTrans}) and (\ref{OmegaA}). We obtain%
\begin{equation}
\hat{H}_{\text{C}}=-\int d^{2}p\,J(\boldsymbol{p})\left[ \hat{\boldsymbol{S}}(-%
\boldsymbol{p})\hat{\boldsymbol{S}}(\boldsymbol{p})+\frac{1}{2N}\hat{\rho}(-\boldsymbol{p})%
\hat{\rho}(\boldsymbol{p})\right] ,  \label{ExchaHamilMom}
\end{equation}%
with%
\begin{align}
J(\boldsymbol{p})& =\ell _{B}^{2}\int d^{2}k\,e^{-\ell _{B}^{2}k^{2}/2}V(\boldsymbol{%
k})\exp \left( -i\ell _{B}^{2}\boldsymbol{p}\wedge \boldsymbol{k}\right)   \notag \\
& =\frac{e^{2}\sqrt{2\pi }\ell _{B}}{4\varepsilon }I_{0}(\ell _{B}^{2}%
\boldsymbol{p}^{2}/4)e^{-\ell _{B}^{2}\boldsymbol{p}^{2}/4}.
\end{align}%
We express the Hamiltonian in the coordinate space. 
\begin{equation}
\hat{H}_{\text{C}}=-\int d^{2}xd^{2}y\,V_{\text{X}}(\boldsymbol{x}-\boldsymbol{y})%
\left[ \hat{\boldsymbol{S}}(\boldsymbol{x})\hat{\boldsymbol{S}}(\boldsymbol{y})+\frac{1}{2N}%
\hat{\rho}(\boldsymbol{x})\hat{\rho}(\boldsymbol{y})\right] ,  \label{ExchaHamilX}
\end{equation}%
where 
\begin{align}
V_{\text{X}}(\boldsymbol{x})& =\frac{e^{2}\ell _{B}}{8\pi \sqrt{2\pi }%
\varepsilon }\int d^{2}p\,e^{i\boldsymbol{px}}e^{-\ell _{B}^{2}\boldsymbol{p}%
^{2}/4}I_{0}(\ell _{B}^{2}\boldsymbol{p}^{2}/4)  \notag \\
& =V(\boldsymbol{x})e^{-\boldsymbol{x}^{2}/2\ell _{B}^{2}}.
\end{align}%
It exhibits the short-range property characteristic to the exchange Coulomb
interaction.

It is worthwhile to mention that we are unable to write the Hamiltonian (\ref%
{ExchaHamilX}) in terms of the projected densities $\bar{\rho}(\boldsymbol{x})$
and $\bar{S}^{A}(\boldsymbol{x})$. If we dare to do so, we would obtain%
\begin{equation}
\hat{H}_{\text{C}}=-\int d^{2}xd^{2}y\,\bar{V}_{\text{X}}(\boldsymbol{x}-\boldsymbol{%
y})\left[ \boldsymbol{\bar{S}}(\boldsymbol{x})\boldsymbol{\bar{S}}(\boldsymbol{y})+\frac{1}{%
2N}\bar{\rho}(\boldsymbol{x})\bar{\rho}(\boldsymbol{y})\right] ,
\label{ExchaHamilXX}
\end{equation}%
with%
\begin{equation}
\bar{V}_{\text{X}}(\boldsymbol{x})=\frac{e^{2}\ell _{B}}{8\pi \sqrt{2\pi }%
\varepsilon }\int d^{2}p\,e^{i\boldsymbol{px}}e^{\ell _{B}^{2}\boldsymbol{p}%
^{2}/4}I_{0}(\ell _{B}^{2}\boldsymbol{p}^{2}/4).
\end{equation}%
However, this is divergent partially due to the factor $e^{\ell _{B}^{2}%
\boldsymbol{p}^{2}/4}$. Thus, it is necessary to use the Weyl-ordered density
operators rather than the projected ones to describe physics in the LLL.

\section{Effective Hamiltonian}

We have derived two expressions (\ref{DirecHamilX}) and (\ref{ExchaHamilX})
for the same Hamiltonian. They are equivalent but the physical picture is
very different. The potential $V_{\text{D}}(\boldsymbol{x})$ in (\ref%
{DirecHamilX})\ is long-ranged, while $V_{\text{X}}(\boldsymbol{x})$ in (\ref%
{ExchaHamilX}) is short-ranged.

In this paper we analyze physics in long-distance scale. The long-distance
limit corresponds to the limit $\ell _{B}\rightarrow 0$. We may replace the
densities with the corresponding ones in the commutative limit, $\hat{\rho}(%
\boldsymbol{x})\rightarrow \rho (\boldsymbol{x})$ and $\hat{\boldsymbol{S}}(\boldsymbol{x}%
)\rightarrow \boldsymbol{S}(\boldsymbol{x})$, with $\rho (\boldsymbol{x})$ and $\boldsymbol{S%
}(\boldsymbol{x})$ being the commutative fields. The Hamiltonian (\ref%
{DirecHamilX}) yields, $\hat{H}_{\text{C}}\rightarrow H_{\text{D}}^{\text{eff%
}}$, where%
\begin{equation}
H_{\text{D}}^{\text{eff}}={\frac{1}{2}}\int d^{2}xd^{2}y\,V_{\text{D}}(%
\boldsymbol{x}-\boldsymbol{y})\rho (\boldsymbol{x})\rho (\boldsymbol{y}),
\label{DirecHamilL}
\end{equation}%
with $\rho (\boldsymbol{x})$ being the commutative field. It represents the
direct Coulomb interaction. On the other hand, the Hamiltonian (\ref%
{ExchaHamilX}) yield, $\hat{H}_{\text{C}}\rightarrow H_{\text{X}}^{\text{eff}%
}$. We make the derivative expansion for a smooth configuration. The first
nontrivial term is 
\begin{equation}
H_{\text{X}}^{\text{eff}}=\frac{2J_{s}}{\rho _{0}^{2}}\int d^{2}x\left[
\partial _{k}\boldsymbol{S}(\boldsymbol{x})\partial _{k}\boldsymbol{S}(\boldsymbol{x})+\frac{%
1}{2N}\partial _{k}\rho (\boldsymbol{x})\partial _{k}\rho (\boldsymbol{x})\right] ,
\label{ExchaHamilL}
\end{equation}%
where $\rho _{0}$ is the electron density and $J_{s}$ is the stiffness
parameter defined by%
\begin{equation}
J_{s}=\frac{\nu ^{2}}{16\sqrt{2\pi }}\frac{e^{2}}{4\pi \varepsilon \ell _{B}}%
.  \label{SpinStiff}
\end{equation}%
We have used the relation $\nu =2\pi \ell _{B}^{2}\rho _{0}$ for the filling
factor. It describes the exchange Coulomb interaction\cite{EzawaX02B}. The
stiffness (\ref{SpinStiff}) agrees with the previous result\cite%
{Moon95B,Ezawa97B}.

The two Hamiltonians (\ref{DirecHamilX}) and (\ref{ExchaHamilX}) are
equivalent when all terms are included. It is intriguing, however, that they
have yielded different effective Hamiltonians in the commutative limit. They
describe entirely different physical effects and they are complementary.
Taking the direct and exchange interactions we construct the full effective
theory\cite{Moon95B,Ezawa97B,EzawaX02B}, 
\begin{equation}
H^{\text{eff}}=H_{\text{D}}^{\text{eff}}+H_{\text{X}}^{\text{eff}}
\label{TotalEffecHamil}
\end{equation}%
with (\ref{DirecHamilL}) and (\ref{ExchaHamilL}).

In the rest of this section we explain why we take the sum (\ref%
{TotalEffecHamil}) as the effective Hamiltonian. In previous sections we
have worked in the Fock representation (\ref{LandaSite}) of the
noncommutativity (\ref{NonCommuXY}). As far as we make an exact analysis the
results are independent of the choice of representation. However, to derive
an effective theory, it is necessary to make a judicious choice to reveal
the essence of the approximation. For the present purpose it is convenient
to adopt the von Neumann lattice representation\cite{EzawaX02B} of the
noncommutativity (\ref{NonCommuXY}), where the Landau-site index $n$ runs
over a lattice with the lattice point being the center of the cyclotron
motion.

We introduce an eigenstate of the annihilation operator $b$ given by (\ref%
{WeylXb}). 
\begin{equation}
b|\beta \rangle =\beta |\beta \rangle .  \label{NeumaA}
\end{equation}%
It is a coherent state by definition,%
\begin{equation}
|\beta \rangle \equiv e^{\beta b^{\dagger }-\beta ^{\ast }b}|0\rangle
=e^{-|\beta |^{2}/2}e^{\beta b^{\dagger }}|0\rangle ,  \label{NeumaC}
\end{equation}%
where $|0\rangle $ is the Fock vacuum obeying $b|0\rangle =0$. The wave
function $\mathfrak{S}_{\beta }(\boldsymbol{x})=\langle \boldsymbol{x}|\beta \rangle 
$ is calculated as%
\begin{equation}
\mathfrak{S}_{\beta }(\boldsymbol{x})={\frac{1}{\sqrt{2\pi \ell _{B}^{2}}}}\exp %
\biggl(-\frac{|z-\sqrt{2}\ell _{B}\beta |^{2}}{4\ell _{B}^{2}}+{\frac{%
i(y\beta _{\Re }-x\beta _{\Im })}{\sqrt{2}\ell _{B}}}\biggr),
\label{NeumaWaveExpli}
\end{equation}%
where $\beta =\beta _{\Re }-i\beta _{\Im }$. It describes an electron
localized around the point $z=\sqrt{2}\ell _{B}\beta $.

The coherent state has the minimum uncertainty subject to the Heisenberg
uncertainty associated with the noncommutativity (\ref{NonCommuXY}). The
state $|\beta \rangle $ corresponds to the classical state describing a
cyclotron motion around the point%
\begin{equation}
x_{\beta }=\sqrt{2}\ell _{B}\beta _{\Re },\quad \quad y_{\beta }=\sqrt{2}%
\ell _{B}\beta _{\Im },
\end{equation}%
as follows from (\ref{WeylXb}) and (\ref{NeumaA}). Since each electron
occupies an area $2\pi \ell _{B}^{2}$, it is reasonable to choose a lattice
with the unit cell area $2\pi \ell _{B}^{2}$. Then, there is one to one
correspondence between the magnetic flux quantum and the lattice site. Such
a lattice is nothing but a von Neumann lattice\cite%
{vonNeumann55,Perelomov71,Bargmann71,Boon81}. The states on a von Neumann
lattice form a minimum complete set\cite{Perelomov71,Bargmann71} in the
lowest Landau level. Thus, we may expand the electron field in terms of
coherent states $\langle \boldsymbol{x}|\beta _{n}\rangle $ as in (\ref%
{FieldExpan}), where $n$ runs over all lattice points.

The merit of this representation is that the wave function $\langle \boldsymbol{x%
}|\beta _{n}\rangle $ is nonvanishing only in a tiny region around the
lattice point $\boldsymbol{x}_{\beta }$ in the limit $\ell _{B}\rightarrow 0$.
The projected density (\ref{RhoStepA}) is well approximated by%
\begin{equation}
\bar{\rho}(\boldsymbol{x})\simeq \sum_{n}\langle \beta _{n}|\boldsymbol{x\rangle }%
\langle \boldsymbol{x}|\beta _{n}\rangle \rho (\beta _{n},\beta _{n}).
\end{equation}%
Consequently, the Weyl-ordered densities (\ref{ProjeDensiP}) and (\ref%
{ProjeDensiS}) are well approximated by 
\begin{align}
\hat{\rho}(\boldsymbol{q})& \simeq \frac{1}{2\pi }\sum_{n}\langle \beta
_{n}|e^{-i\boldsymbol{qX}}|\beta _{n}\rangle \rho (\beta _{n},\beta _{n}), \\
\hat{S}^{A}(\boldsymbol{q})& \simeq \frac{1}{2\pi }\sum_{n}\langle \beta
_{n}|e^{-i\boldsymbol{qX}}|\beta _{n}\rangle S^{A}(\beta _{n},\beta _{n}).
\end{align}%
The main contribution to $\hat{\rho}(\boldsymbol{x})$ and $\hat{S}^{A}(\boldsymbol{x}%
)$ come from the electrons in one Landau site $|\beta \rangle $ containing
the position $\boldsymbol{x}$. With this approximation $\hat{\rho}$ and $\hat{S}%
^{A}$ satisfy the U(N) algebra rather than the W$_{\infty }$(N) algebra.
Hence, they correspond to the densities in the commutative limit.

We now examine the Coulomb energy (\ref{HamilInC}), or%
\begin{equation}
\hat{H}_{\text{C}}=\sum_{mnij}\sum_{\sigma \tau }{V}_{mnij}c_{\sigma
}^{\dagger }(\beta _{m})c_{\tau }^{\dagger }(\beta _{i})c_{\tau }(\beta
_{j})c_{\sigma }(\beta _{n}),  \label{HamilInCx}
\end{equation}%
where the indices $m,n,i,j$ run over the lattice points. In a semiclassical
approximation the matrix element matters. It vanishes unless $\beta
_{m}=\beta _{n}$ and $\beta _{i}=\beta _{j},$ or $\beta _{m}=\beta _{j}$ and 
$\beta _{i}=\beta _{n}$. These two terms represent the direct and exchange
Coulomb interactions, respectively, which are the dominant ones in (\ref%
{HamilInCx}). We may summarize them as\cite{EzawaX02B},%
\begin{equation}
\hat{H}_{\text{D}}=\sum_{mi}{V}_{mmii}\rho (\beta _{m},\beta _{m})\rho
(\beta _{i},\beta _{i}),  \label{DirecHamil}
\end{equation}%
and%
\begin{align}
\hat{H}_{\text{X}} &=-2\sum_{mi}{V}_{miim}[S^{A}(\beta _{m},\beta
_{m})S^{A}(\beta _{i},\beta _{i})  \notag \\
&\hspace{0.5in}+\frac{1}{2N}\rho (\beta _{m},\beta _{m})\rho (\beta
_{i},\beta _{i})],  \label{ExchaHamil}
\end{align}%
which have no parts in common by construction. They are the special parts of
the two equivalent and exact Hamiltonians (\ref{EquivHamilD}) and (\ref%
{EquivHamilX}). Furthermore, it is clear from our arguments that they are
reduced to (\ref{DirecHamil}) and (\ref{ExchaHamil}) in the commutative
limit ($\ell _{B}\rightarrow 0$). Hence, we take (\ref{TotalEffecHamil}) as
the effective Hamiltonian in the commutative limit.

\section{Goldstone Modes}

It is convenient to study quantum coherence based on the Hamiltonian (\ref%
{ExchaHamilMom}). It is minimized by the uniform configuration of the
isospin as well as the density,%
\begin{equation}
\hat{\rho}(\boldsymbol{p})=2\pi \rho _{0}\delta (\boldsymbol{p}),\quad \hat{\boldsymbol{S%
}}(\boldsymbol{p})=2\pi \boldsymbol{S}_{0}\delta (\boldsymbol{p}).
\end{equation}%
Consequently, all isospins are spontaneously polarized into one isospin
direction. In the zero-momentum sector the W$_{\infty}$(N) algebra (\ref%
{WAlgeb}) is reduced to the U(N)\ algebra, 
\begin{subequations}
\label{DensiAlgebZero}
\begin{equation}
\lbrack \hat{\rho}_{0},\hat{\rho}_{0}]={0},\quad \lbrack \hat{S}_{0}^{A},%
\hat{\rho}_{0}]={0},\quad \lbrack \hat{S}_{0}^{A},\hat{S}_{0}^{B}]={\frac{i}{%
2\pi }}f^{ABC}\hat{S}_{0}^{C},
\end{equation}%
where $\hat{\rho}_{0}=\hat{\rho}(\boldsymbol{p}=0)$ and $\hat{S}_{0}^{A}=\hat{S}%
^{A}(\boldsymbol{p}=0)$.

The ground state is characterized by the algebra U(N) rather than W$_{\infty
}$(N). At $\nu =1$, there are $N$ degenerate isospin states any one of which
may be spontaneously filled up to make a ground state. At $\nu =k$, $k$ of
the $N$ degenerate states are occupied and $(N-k)$ of them are empty. Hence,
the unbroken global symmetry is SU(k)$\otimes $SU(N-k)$\otimes $U(1),
implying a spontaneous breaking of the global SU(N) symmetry, 
\end{subequations}
\begin{equation}
\text{SU(N)}\rightarrow \text{SU(k)}\otimes \text{SU(N-k)}\otimes \text{U(1)}%
.
\end{equation}%
The target space is the coset space,%
\begin{equation}
G_{N,k}=\text{SU(N)}/[\text{SU(k)}\otimes \text{SU(N-k)}\otimes \text{U(1)}],
\label{GrassNk}
\end{equation}%
which is known as the Grassmannian $G_{N,k}$ manifold. Its real dimension is 
$N^{2}-k^{2}-(N-k)^{2}=2k(N-k)$. We expect $k(N-k)$ complex Goldstone modes
to appear as a result of this spontaneous symmetry breaking. Note that $%
G_{N,k}=G_{N,N-k}$ as a manifold. Hence, the physics at $\nu =k$ and $\nu
=N-k$ is identical. It is enough to study the case for $\nu \leq N/2$.

We analyze the Goldstone modes based on the effective Hamiltonian (\ref%
{TotalEffecHamil}). Because the QH system is incompressible\cite%
{BookEzawa,BookDasSarma}, we may set $\rho (\boldsymbol{x})=\rho _{0}$ as far as
perturbational fluctuations are concerned. When we define the normalized
isospin field $\mathscr{S}^{A}(\boldsymbol{x})$\ by 
\begin{equation}
S^{A}(\boldsymbol{x})=\rho (\boldsymbol{x})\mathscr{S}^{A}(\boldsymbol{x}),
\label{SpinNoSpin}
\end{equation}%
the Hamiltonian (\ref{TotalEffecHamil}) yields%
\begin{equation}
H^{\text{eff}}=2J_{s}\sum_{A}\int d^{2}x[\partial _{k}\mathscr{S}^{A}(\boldsymbol{%
x})]^{2},  \label{ExchaSU4}
\end{equation}%
up to the leading order in the derivative expansion. This is the SU(N)
nonlinear sigma model.

We first study the filling factor $\nu =1$. It is convenient to use the
composite boson (CB) theory of quantum Hall ferromagnets\cite{Ezawa99L} to
identify the dynamic degree of freedom. The CB field $\phi ^{\sigma }(%
\boldsymbol{x})$ is defined by making a singular phase transformation\cite%
{Girvin87L} to the electron field $\psi ^{\sigma }(\boldsymbol{x})$, 
\begin{equation}
\phi ^{\sigma }(\boldsymbol{x})=e^{-ie\Theta (\boldsymbol{x})}\psi ^{\sigma }(%
\boldsymbol{x}),  \label{BareCB}
\end{equation}%
where the phase field $\Theta (\boldsymbol{x})$ attaches one flux quantum to
each electron via the relation,%
\begin{equation}
\varepsilon _{ij}\partial _{i}\partial _{j}\Theta (\boldsymbol{x})=\Phi _{\text{D%
}}\rho (\boldsymbol{x}).  \label{PhaseField}
\end{equation}%
We then introduce the normalized CB field $n^{\sigma }(\boldsymbol{x})$ by 
\begin{equation}
\phi ^{\sigma }(\boldsymbol{x})=\phi (\boldsymbol{x})n^{\sigma }(\boldsymbol{x}),
\label{NormaCB}
\end{equation}%
where the $N$-component field $n^{\sigma }(\boldsymbol{x})$ obeys the constraint 
$\sum_{\sigma }n^{\sigma \dagger }(\boldsymbol{x})n^{\sigma }(\boldsymbol{x})=1$:
Such a field is the $CP^{N-1}$ field\cite{DAdda78NPB}. On the other hand, $%
\phi ^{\ast }(\boldsymbol{x})\phi (\boldsymbol{x})=\rho (\boldsymbol{x)}$ for the U(1)
field $\phi (\boldsymbol{x})$.

Formula (\ref{NormaCB}) is interpreted as a charge-isospin separation.
Indeed, by substituting (\ref{BareCB}) together with (\ref{NormaCB}) into
the kinetic Hamiltonian (\ref{HamilKinem}), the electromagnetic field $A_{k}(%
\boldsymbol{x})$ is found to be coupled only with the U(1) field $\phi (\boldsymbol{x%
})$. Thus, the charge is carried by $\phi (\boldsymbol{x})$, while the isospin
is carried by $n^{\sigma }(\boldsymbol{x})$.

In terms of the $CP^{N-1}$ field $\boldsymbol{n}(\boldsymbol{x})$, the isospin $%
\mathscr{S}^{A}(\boldsymbol{x})$\ field reads 
\begin{equation}
\mathscr{S}^{A}(\boldsymbol{x})={\frac{1}{2}}\boldsymbol{n}^{\dagger }(\boldsymbol{x}%
)\lambda ^{A}\boldsymbol{n}(\boldsymbol{x}),  \label{IpinDensi}
\end{equation}%
with which the Hamiltonian (\ref{ExchaSU4}) is equivalent to 
\begin{equation}
H^{\text{eff}}=2J_{s}\int d^{2}x(\partial _{j}\boldsymbol{n}^{\dagger }+iK_{j}%
\boldsymbol{n}^{\dagger })\!\cdot \!(\partial _{j}\boldsymbol{n}\boldsymbol{\ }-iK_{j}%
\boldsymbol{n}),  \label{ExchaCPx}
\end{equation}%
where $K_{\mu }(\boldsymbol{x})=-i\boldsymbol{n}^{\dagger }(\boldsymbol{x})\partial
_{\mu }\boldsymbol{n}(\boldsymbol{x})$. The field $K_{\mu }$ is not a dynamic field
because of the absence of the kinetic term. The $N$-component field $\boldsymbol{%
n}(\boldsymbol{x})$ has $N-1$ independent complex components: They are the
Goldstone modes.

There are $N$ degenerate states any one of which can be chosen as the ground
state. For definiteness, let us choose%
\begin{equation}
\boldsymbol{n}_{\text{g}}(\boldsymbol{x})=(1,0,\ldots ,0)  \label{GrounCP}
\end{equation}%
as a ground state. The $CP^{N-1}$ field is parametrized as%
\begin{equation}
\boldsymbol{n}(\boldsymbol{x})=(1,\eta _{1},\ldots ,\eta _{N-1})
\end{equation}%
up to the lowest order of perturbation, where $\eta _{i}$ are the $N-1$
Goldstone modes.

We next study the case $\nu =k$. To describe $k$ electrons in one Landau
site we introduce $k$ normalized CB fields $n_{i}^{\sigma }(\boldsymbol{x})$.
They should be orthogonal one to another, 
\begin{equation}
\boldsymbol{n}_{i}^{\dagger }(\boldsymbol{x})\cdot \boldsymbol{n}_{j}(\boldsymbol{x})=\delta
_{ij},
\end{equation}%
because they are not ordinary bosons but hard-core bosons. (Two hard-core
bosons never occupy a single quantum state just like electrons subject to
the Pauli exclusion principle.) We then construct an $N\times k$ matrix
field 
\begin{equation}
Z(\boldsymbol{x})=(\boldsymbol{n}_{1},\boldsymbol{n}_{2},\cdots ,\boldsymbol{n}_{N}),
\label{G42Field}
\end{equation}%
using a set of $k$ fields subject to this normalization condition, or 
\begin{equation}
Z^{\dagger }Z=1.
\end{equation}%
Though we have introduced $k$ fields $\boldsymbol{n}_{i}(\boldsymbol{x})$, we cannot
distinguish them quantum mechanically since they describe $k$ identical
electrons in the same Landau site. Namely, two fields $Z(\boldsymbol{x})$ and $%
Z^{\prime }(\boldsymbol{x})$ are indistinguishable physically when they are
related by a local U(N) transformation $U(\boldsymbol{x})$,%
\begin{equation}
Z^{\prime }(\boldsymbol{x})=Z(\boldsymbol{x})U(\boldsymbol{x}).
\end{equation}%
By identifying these two fields $Z(\boldsymbol{x})$ and $Z^{\prime }(\boldsymbol{x})$%
, the $N\times k$ matrix field $Z(\boldsymbol{x})$ takes values on the
Grassmannian manifold $G_{N,k}$ defined by (\ref{GrassNk}). The field $Z(%
\boldsymbol{x})$ is no longer a set of $k$ independent $CP^{N-1}$ fields. It is
a new object, called the Grassmannian field, carrying $k(N-k)$ complex
degrees of freedom.

At $\nu =k$ the isospin field $\boldsymbol{\mathscr{S}}(\boldsymbol{x})$ is
represented in terms of the Grassmannian $G_{N,k}$ field $Z(\boldsymbol{x})$\ as 
\begin{equation}
\mathscr{S}^{A}(\boldsymbol{x})=\text{Tr}\left[ Z^{\dagger }(\boldsymbol{x}){\frac{%
\lambda ^{A}}{2}}Z(\boldsymbol{x})\right] =\frac{1}{2}\sum_{i}\boldsymbol{n}%
_{i}^{\dagger }(\boldsymbol{x})\lambda ^{A}\boldsymbol{n}_{i}(\boldsymbol{x}).
\end{equation}%
It is a simple sum of isospins of $k$ electrons in one Landau site. With
this identification we are able to rewrite the SU(N) sigma-model Hamiltonian
(\ref{ExchaSU4}) as 
\begin{equation}
\mathcal{H}^{\text{eff}}=2J_{s}\;\text{Tr}\left[ (\partial
_{j}Z-iZK_{j})^{\dagger })(\partial _{j}Z-iZK_{j})\right] ,
\label{ExchaGrass}
\end{equation}%
where 
\begin{equation}
K_{\mu }(\boldsymbol{x})=-iZ^{\dagger }(\boldsymbol{x})\partial _{\mu }Z(\boldsymbol{x}).
\label{AuxiaK}
\end{equation}%
This Hamiltonian is known as the Grassmannian sigma-model Hamiltonian\cite%
{MacFarlane82PLB}. It has the local U(N) gauge symmetry, 
\begin{align}
Z(\boldsymbol{x})& \rightarrow Z(\boldsymbol{x})U(\boldsymbol{x}),  \notag \\
K_{\mu }(\boldsymbol{x})& \rightarrow U(\boldsymbol{x})^{\dagger }K_{\mu }(\boldsymbol{x}%
)U(\boldsymbol{x})-iU(\boldsymbol{x})^{\dagger }\partial _{\mu }U(\boldsymbol{x}).
\label{LocalU2}
\end{align}%
The gauge field $K_{\mu }$ is not a dynamic field because of the absence of
the kinetic term.

The $G_{N,k}$ field has $k(N-k)$ independent complex components: They are
the Goldstone modes $\eta _{ij}$\ parametrized as 
\begin{equation}
Z=\left( 
\begin{array}{cccc}
1 & 0 & \cdots & 0 \\ 
0 & 1 & \cdots & 0 \\ 
\vdots & \vdots & \ddots & \vdots \\ 
0 & 0 & \cdots & 1 \\ 
\eta _{1,1} & \eta _{2,1} & \cdots & \eta _{k,1} \\ 
\vdots & \vdots & \vdots & \vdots \\ 
\eta _{1,N-k} & \eta _{2,N-k} & \cdots & \eta _{k,N-k}%
\end{array}%
\right) U(\boldsymbol{x}),  \label{G42Golds}
\end{equation}%
up to the lowest order of perturbation. The corresponding ground state $Z_{%
\text{g}}$\ is given by setting $\eta _{ij}=0$ in (\ref{G42Golds}). We make
a gauge choice such that $U(\boldsymbol{x})=1$\ in (\ref{G42Golds}).
Substituting the parametrization (\ref{G42Golds}) of the Grassmannian field
into the Hamiltonian (\ref{ExchaGrass}), we expand it up to the second
order, 
\begin{align}
\mathcal{H}^{\text{eff}} &={2}J_{s}\sum_{s=1}^{k(N-k)}\partial _{k}\eta
_{s}^{\dagger }(\boldsymbol{x})\partial _{k}\eta _{s}(\boldsymbol{x})  \notag \\
&={\frac{J_{s}}{2}}\sum_{s=1}^{k(N-k)}\bigl\{(\partial _{k}\sigma
_{s})^{2}+(\partial _{k}\vartheta _{s})^{2}\bigr\},  \label{EffecHamilSpin}
\end{align}%
where $\eta _{s}(\boldsymbol{x})$ stands for $\eta _{ij}(\boldsymbol{x})$ and 
\begin{equation}
\eta _{s}(\boldsymbol{x})={\frac{1}{2}}\bigl(\sigma _{s}(\boldsymbol{x})+i\vartheta
_{s}(\boldsymbol{x})\bigr)  \label{ZetaField}
\end{equation}%
with%
\begin{equation}
\lbrack \sigma _{s}(\boldsymbol{x}),\vartheta _{t}(\boldsymbol{y})]=2i\rho _{\Phi
}^{-1}\delta _{st}\delta (\boldsymbol{x}-\boldsymbol{y}).
\end{equation}%
Here, $\rho _{\Phi }=\frac{1}{k}\rho _{0}=1/(2\pi \ell _{B}^{2})$ is the
magnetic flux density, that is, the density of Landau sites. This
Hamiltonian realizes the SU(N) symmetry nonlinearly.

\section{Grassmannian G$_{N,k}$ Solitons}

The existence of topological solitons, which we call $G_{N,k}$ solitons, is
guaranteed by the homotopy theorem 
\begin{equation}
\pi _{2}(G_{N,k})=\pi _{1}(\text{U(1)})=\mathbb{Z},  \label{HomotTheor}
\end{equation}%
which follows from (\ref{GrassNk}), where we have used $\pi _{2}(G/H)=\pi
_{1}(H)$ (when G is simply connected) and $\pi _{n}(G\otimes G^{\prime
})=\pi _{n}(G)\oplus \pi _{n}(G^{\prime })$. The topological charge is
defined\cite{MacFarlane82PLB} as a gauge invariant by 
\begin{equation}
Q={\frac{i}{2\pi }}\int d^{2}x\,\epsilon _{jk}\text{Tr}\left[ (\partial
_{j}Z-iZK_{j})^{\dagger }(\partial _{k}Z-iZK_{k})\right] .
\label{TopolCharg}
\end{equation}%
It is a topological invariant since it is the charge of the topological
current, $Q=\int d^{2}xJ_{\text{sol}}^{0}(\boldsymbol{x})$, with 
\begin{equation}
J_{\text{sol}}^{\mu }(\boldsymbol{x})={\frac{i}{2\pi }}\epsilon ^{\mu \nu
\lambda }\text{Tr}\left[ (\partial _{\nu }Z)^{\dagger }(\partial _{\lambda
}Z)\right] .
\end{equation}%
Based on (\ref{G42Field}) we rewrite it as%
\begin{equation}
J_{\text{sol}}^{\mu }(\boldsymbol{x})={\frac{i}{2\pi }}\sum_{i=1}^{k}\epsilon
^{\mu \nu \lambda }(\partial _{\nu }\boldsymbol{n}_{i})^{\dagger }\cdot
(\partial _{\lambda }\boldsymbol{n}_{i}).  \label{TopolChargKN}
\end{equation}%
It is the sum of the topological charges associated with the $k$ $CP^{N-1}$
fields $\boldsymbol{n}_{i}$. Hence, the $G_{N,k}$ soliton consists of $k$ $%
CP^{N-1}$ solitons,%
\begin{equation}
n_{i}^{\sigma }(\boldsymbol{x})=\frac{1}{\sqrt{\sum_{\tau =1}^{N}\omega
_{i}^{\tau }(z)|}}\omega _{i}^{\sigma }(z),  \label{EachCP}
\end{equation}%
where $\omega _{i}^{\sigma }(z)$ are arbitrary analytic functions.

Grassmannian solitons are constructed as classical configurations, as
dictated by the homotopy theorem (\ref{HomotTheor}). They are BPS states of
the Grassmannian sigma model (\ref{ExchaGrass}). Indeed, the following
inequality holds\cite{MacFarlane82PLB} between the exchange energy (\ref%
{ExchaGrass}) and the topological charge (\ref{TopolCharg}),%
\begin{equation}
H^{\text{eff}}\geqq 4\pi J_{s}Q,  \label{CondiBPS}
\end{equation}%
where the equality is achieved by the $G_{N,k}$ soliton.

The simplest soliton would be a set of one $CP^{3}$ soliton in one component
and the ground state in all others. An example reads 
\begin{equation}
Z^{1}=\frac{1}{\sqrt{|z|^{2}+\kappa ^{2}}}\left( 
\begin{array}{cccc}
z & 0 & \cdots & 0 \\ 
0 & \sqrt{|z|^{2}+\kappa ^{2}} & \cdots & 0 \\ 
\vdots & \vdots & \ddots & \vdots \\ 
0 & 0 & \cdots & \sqrt{|z|^{2}+\kappa ^{2}} \\ 
\kappa & 0 & \cdots & 0 \\ 
\vdots & \vdots & \vdots & \vdots \\ 
0 & 0 & \cdots & 0%
\end{array}%
\right) ,  \label{G42Sky1}
\end{equation}%
for which the topological charge (\ref{TopolCharg}) is $Q=1$. We argue in
the next section that the simplest $G_{N,k}$ soliton (\ref{G42Sky1}) is
ruled out since it is not confined to the LLL. As we shall see, the simplest
allowed one is a set of $k$ $CP^{N-1}$ solitons with $Q=k$\ when $k\leq 
\frac{1}{2}N$. We give an instance of a $G_{5,2}$ soliton,%
\begin{equation*}
Z^{2}=\frac{1}{\sqrt{|z|^{2}+\kappa ^{2}}}\left( 
\begin{array}{cc}
z & 0 \\ 
0 & z \\ 
\kappa & 0 \\ 
0 & \kappa \\ 
0 & 0%
\end{array}%
\right) ,
\end{equation*}%
for which $Q=2$.

\section{Charge-Isospin Relation}

\label{SecQuasi}

We have found topological solitons in the effective Hamiltonian (\ref%
{ExchaSU4}). However, it determines only the isospin part of the excitation
in the charge-isospin separation formula (\ref{NormaCB}). It is necessary to
analyze how the isospin modulation affects the charge part. This can be done
by requiring the LLL condition (\ref{LLLcondi}) on soliton states with a
combined charge-isospin modulation. Recall that the effective Hamiltonian is
an ordinary local field theory, in which we have made the charge-isospin
separation. Hence, it is necessary to examine whether these solitons are
confined to the LLL.

The LLL condition becomes particularly simple in the dressed-CB picture\cite%
{BookEzawa} in the symmetric gauge with the angular-momentum state for the
Landau site. We define the dressed-CB field $\varphi ^{\sigma }(\boldsymbol{x})$
by\cite{Ezawa99L,Read89L}%
\begin{equation}
\varphi ^{\sigma }(\boldsymbol{x})=e^{-\mathcal{A}(\boldsymbol{x})}\phi ^{\sigma }(%
\boldsymbol{x})=e^{-\mathcal{A}(\boldsymbol{x})}\sqrt{\rho (\boldsymbol{x})}n^{\sigma }(%
\boldsymbol{x}),  \label{DressCB}
\end{equation}%
where $\phi ^{\sigma }(\boldsymbol{x})$ is the CB field (\ref{BareCB}) with the
U(1) phase factor removed; the auxiliary field ${\mathcal{A}}(\boldsymbol{x})$
is determined by%
\begin{equation}
\boldsymbol{\nabla }^{2}{\mathcal{A}}(\boldsymbol{x})=2\pi \left( \rho (\boldsymbol{x}%
)-\rho _{0}\right) ,  \label{Dress2OperaE}
\end{equation}%
as follows from the condition (\ref{PhaseField}) on the phase field. It is
straightforward to rewrite the LLL condition (\ref{LLLcondi}) as 
\begin{equation}
\frac{\partial }{\partial z^{\ast }}\varphi ^{\sigma }(\boldsymbol{x})|\mathfrak{%
S}\rangle =0.  \label{CondiLLL}
\end{equation}%
We take a coherent state of $\varphi ^{\sigma }(\boldsymbol{x})$, for which (\ref%
{CondiLLL}) implies%
\begin{equation}
\varphi ^{\sigma }(\boldsymbol{x})|\mathfrak{S}\rangle =\omega ^{\sigma }(z)|%
\mathfrak{S}\rangle ,
\end{equation}%
where $\omega ^{\sigma }(z)$ is an analytic function. The coherent state $|%
\mathfrak{S}\rangle $ must be an eigenstate of the density operator $\rho (%
\boldsymbol{x})$ and a coherent state of the $CP^{3}$ field $\boldsymbol{n}(\boldsymbol{x%
})$ since they commute with each other. Hence we have 
\begin{equation}
e^{-\mathcal{A}^{\text{cl}}(\boldsymbol{x})}\sqrt{\rho ^{\text{cl}}(\boldsymbol{x})}%
n^{\text{cl(}\sigma )}(\boldsymbol{x})=\omega ^{\sigma }(z),  \label{Dress2CB}
\end{equation}%
where $\mathcal{A}^{\text{cl}}(\boldsymbol{x})$, $\rho ^{\text{cl}}(\boldsymbol{x})$
and $n^{\text{cl(}\sigma )}(\boldsymbol{x})$ are classical fields. The
holomorphicity of $\omega ^{\sigma }(z)$ is a consequence of the requirement
that the excitation is confined to the LLL. This is the LLL condition for
soliton states.

We study the $G_{N,k}$ soliton at $\nu =k$. When the CB field acts on the
state at $\nu =k$, it picks up contributions from $k$ electrons at each
point, 
\begin{equation}
\boldsymbol{n}(\boldsymbol{x})|\Phi \rangle =\boldsymbol{n}^{\text{cl}}(\boldsymbol{x})|\Phi
\rangle =\sum_{i=1}^{k}\boldsymbol{n}_{i}^{\text{cl}}(\boldsymbol{x})|\Phi \rangle ,
\label{TwoCP}
\end{equation}%
together with $\boldsymbol{n}_{i}^{\text{cl}}(\boldsymbol{x})\cdot \boldsymbol{n}_{j}^{%
\text{cl}}(\boldsymbol{x})=\delta _{ij}$ and 
\begin{equation}
\rho (\boldsymbol{x})|\Phi \rangle =\rho ^{\text{cl}}(\boldsymbol{x})|\Phi \rangle
=\sum_{i=1}^{k}\rho _{i}^{\text{cl}}(\boldsymbol{x})|\Phi \rangle .
\label{SolitTotalDensi}
\end{equation}%
We may solve (\ref{Dress2CB}) as 
\begin{equation}
n^{\text{cl(}\sigma )}(\boldsymbol{x})=\frac{\sqrt{k}}{\sqrt{\sum_{\tau
=1}^{N}|\omega ^{\tau }(x)|^{2}}}\omega ^{\sigma }(z).  \label{CondiCP}
\end{equation}%
From (\ref{EachCP}), (\ref{TwoCP}) and (\ref{CondiCP}) we find $\omega
^{\sigma }(z)=\sum_{i=1}^{k}\omega _{i}^{\sigma }(z)$ and $\sum_{\sigma
}|\omega _{1}^{\sigma }(z)|^{2}=\sum_{\sigma }|\omega _{2}^{\sigma
}(z)|^{2}=\cdots =\sum_{\sigma }|\omega _{k}^{\sigma }(z)|^{2}$. Thus, the
LLL condition (\ref{Dress2CB}) holds for each component,%
\begin{equation}
e^{-\mathcal{A}^{\text{cl}}(\boldsymbol{x})}\sqrt{\rho ^{\text{cl}}(\boldsymbol{x})}%
n_{i}^{\text{cl(}\sigma )}(\boldsymbol{x})=\omega _{i}^{\sigma }(z),
\label{Dress2CBx}
\end{equation}%
where $\rho ^{\text{cl}}(\boldsymbol{x})$ represents the total density (\ref%
{SolitTotalDensi}). Substituting (\ref{EachCP}) into (\ref{Dress2CBx}) and
using (\ref{Dress2OperaE}) we derive the soliton equation 
\begin{equation}
{\frac{1}{4\pi }}\boldsymbol{\nabla }^{2}\ln \rho ^{\text{cl}}(\boldsymbol{x})-\rho
^{\text{cl}}(\boldsymbol{x})+\rho _{0}=j_{\text{sol}}^{0}(\boldsymbol{x}),
\label{SolitEq}
\end{equation}%
where 
\begin{equation}
j_{\text{sol}}^{0}(\boldsymbol{x})={\frac{1}{4\pi }}\boldsymbol{\nabla }^{2}\ln
\sum_{\sigma =1}^{N}|\omega _{1}^{\sigma }(z)|^{2}.  \label{Semi2Charg}
\end{equation}%
It is easy to see that the topological charge density (\ref{TopolChargKN})
is given by $J_{\text{sol}}^{0}(\boldsymbol{x})=kj_{\text{sol}}^{0}(\boldsymbol{x})$.

The soliton equation (\ref{SolitEq}) is solved iteratively. The density
modulation is given by%
\begin{equation}
\delta \rho ^{\text{cl}}(\boldsymbol{x})=\rho ^{\text{cl}}(\boldsymbol{x})-\rho
_{0}=-J_{\text{sol}}^{0}(\boldsymbol{x})+\cdots .
\end{equation}%
The iteration corresponds to the derivative expansion. The leading term is
precisely the topological charge density. The electric charge density
modulation is%
\begin{equation*}
\delta Q_{e}=-e\delta \rho ^{\text{cl}}(\boldsymbol{x}).
\end{equation*}
An isospin rotation turns out to induce the density modulation of the
electric charge according to this formula. The total change of the charge
due to a soliton excitation is given by $\int d^{2}x\,\delta Q_{e}=ke$.

We have found that $k$ $CP^{N-1}$ fields (\ref{EachCP}) have the same
normalization. Namely, among many soliton configurations in the $G_{N,k}$
sigma model, only a special type of configurations are allowed by requiring
the LLL condition. The $G_{N,k}$ soliton has a general expression,%
\begin{equation}
Z^{1}=\frac{1}{\sqrt{\sum_{\sigma }|\omega _{1}^{\sigma }(z)|^{2}}}\left( 
\begin{array}{cccc}
\omega _{1}^{1}(z) & \omega _{2}^{1}(z) & \cdots & \omega _{k}^{1}(z) \\ 
\omega _{1}^{2}(z) & \omega _{2}^{2}(z) & \cdots & \omega _{k}^{2}(z) \\ 
\vdots & \vdots & \ddots & \vdots \\ 
\omega _{1}^{k}(z) & \omega _{2}^{k}(z) & \cdots & \omega _{k}^{k}(z) \\ 
\omega _{1}^{k+1}(z) & \omega _{2}^{k+1}(z) & \cdots & \omega _{k}^{k+1}(z)
\\ 
\vdots & \vdots & \vdots & \vdots \\ 
\omega _{1}^{N}(z) & \omega _{2}^{N}(z) & \cdots & \omega _{k}^{N}(z)%
\end{array}%
\right) .  \label{GenerGrassSolit}
\end{equation}%
This rules out the soliton (\ref{G42Sky1}) with $Q=1$.

The origin of this peculiarity may be attributed to the charge-isospin
separation formula (\ref{NormaCB}), by way of which the normalized CB field $%
\boldsymbol{n}(\boldsymbol{x})$ is introduced. It is essential that the total
electron density $\rho (\boldsymbol{x})$ is common to all the $N$ components.
Otherwise, the symmetry SU(N) is explicitly broken by hand. As a
consequence, even if we try to excite a soliton only in one of the
components, the density modulation associated with it affects equally
electrons in other components to satisfy the LLL condition. It is impossible
to have a soliton excitation only in one of the components.

\section{Applications}

We have studied the dynamics of $N$-component electrons projected to the
LLL. We have shown that the long-distance physics is described by the
Grassmannian sigma model. Physically it is realized by multilayer QH
systems. Various experiments have already been carried out not only on
monolayer QH systems with spin ($N=2$) but also on bilayer QH systems with
spin ($N=4$).

In the monolayer QH system with spin the effective Hamiltonian consists of
the exchange term, the direct term and the Zeeman term. It is well
approximated by\newline
\begin{align}
H^{\text{eff}}& =2J_{s}\sum_{A=x,y,z}\int d^{2}x[\partial _{k}\mathscr{S}^{A}(%
\boldsymbol{x})]^{2}+\rho _{0}\Delta _{\text{Z}}\int d^{2}x\,\mathscr{S}^{z}(%
\boldsymbol{x})  \notag \\
& +{\frac{1}{2}}\int d^{2}xd^{2}y\,V_{\text{D}}(\boldsymbol{x}-\boldsymbol{y})\rho (%
\boldsymbol{x})\rho (\boldsymbol{y}),
\end{align}%
where $\mathscr{S}^{A}(\boldsymbol{x})$ is the spin SU(2) field, and $\Delta _{%
\text{Z}}$ is the energy gap between the one-particle up-spin and down-spin
states due to the Zeeman effect. The direct Coulomb term is necessary since
the soliton modulates the density $\rho (\boldsymbol{x})$ according to the
soliton equation (\ref{SolitEq}). The system possesses one Goldstone mode,
which is made massive by the Zeeman term. The topological soliton is the $%
G_{2,1}$ ($CP^{1}$) soliton, represented by the Grassmannian field (\ref%
{GenerGrassSolit}) or%
\begin{equation}
Z=\frac{1}{\sqrt{|z|^{2}+\kappa ^{2}}}\left( 
\begin{array}{c}
z \\ 
\kappa%
\end{array}%
\right) ,
\end{equation}%
where $\kappa $ represents the size of the soliton. The exchange energy is
independent of it. As $\kappa $ increases, the direct Coulomb energy is
decreased while the Zeeman energy increases. Thus, it is determined to
optimize these two energies\cite{Sondhi93B}.

Let us discuss bilayer QH systems\cite{BookEzawa,EzawaX02B} somewhat in
detail. They are very interesting because they exhibit various novel quantum
coherent phenomena due to the rich isospin degree of freedom. An electron in
bilayer QH systems is labeled by the spin SU(2) and the pseudospin SU(2)
representing the layer degree of freedom. A group that accommodates the spin
SU(2) and pseudospin SU(2) is SU(4), which is a good symmetry of the system
when the two layers are placed close enough. It reminds us of the grand
unified theory (GUT), where the standard-model gauge group SU(3)$\otimes $%
SU(2)$\otimes $U(1) is incorporated into a larger group. However, there
exists a big difference: In QH systems it is a \textit{global} symmetry
which is in problem, while in GUT it is a \textit{local} (gauge) symmetry.
Thus, Goldstone bosons appear in QH systems, while some gauge bosons get
massive by eating Goldstone bosons in GUT.

In bilayer QH systems we introduce $4$-component spinors, and analyze
spontaneous symmetry breaking, Goldstone bosons and topological solitons.
According to our general arguments, at the integer filling factor $\nu =k$,
complex $k(4-k)$ Goldstone bosons appear accompanied with a spontaneous
breakdown of the global SU(4) symmetry. In actual systems the SU(4) symmetry
is broken explicitly but only softly by sufficiently weak Zeeman or
tunneling interactions. All Goldstone modes are made massive by these
interactions. Topological solitons are Grassmannian $G_{4,k}$ solitons. The
topological mapping is determined by the U(1) group, as follows from (\ref%
{HomotTheor}). It reminds us of the U(1) monopole in the context of the GUT,
which appears when the GUT gauge group is broken to a subgroup including
U(1) group.

The integer filling factor is up to $4$ in the LLL. The nontrivial
Grassmannian manifold is realized only at $\nu =1$, $2$ and $3$. We explain
what we expect at these filling factors. At $\nu =1$, the breakdown pattern
of the SU(4) symmetry is%
\begin{equation}
\text{SU(4)}\rightarrow \text{U(1)}\otimes \text{SU(3)}.
\end{equation}%
There arise three complex pseudo-Goldstone modes. The fifteen generators of
SU(4) accommodate \textquotedblleft six\textquotedblright\ different SU(2)
generators. Some internal different SU(2) symmetries also break down when
the SU(4) symmetry breaks down. These three Goldstone modes appear due to
the symmetry breaking of \textquotedblleft three\textquotedblright\ internal
SU(2)'s. Topological excitations are $G_{4,1}$($CP^{3}$) solitons. The mode
associated with the pseudospin SU(2) breaking induces the Josephson-like
tunneling current\cite{Ezawa92IJMPB}, whose signals have been detected
experimentally\cite{Spielman00L}.

At $\nu =2$, the breakdown pattern of the SU(4) symmetry is 
\begin{equation}
\text{SU(4)}\rightarrow \text{U(1)}\otimes \text{SU(2)}\otimes \text{SU(2)}.
\end{equation}%
There arise four complex Goldstone modes. The topological excitations are $%
G_{4,2}$ solitons. In actual samples the degeneracy is resolved by the
Zeeman and tunneling interactions. According to their relative strength we
have two phases, where either the spin or the pseudospin is polarized. In
the spin-polarized phase, one $G_{4,2}$ soliton flips twice as much spins as
one $CP^{3}$ soliton does, whose specific features have already been
observed experimentally\cite{Kumada00L}: See Ref.\cite{Hasebe02B} for more
details.

We have mentioned the property $G_{N,k}=G_{N,N-k}$ of the Grassmannian
manifold. We can explain it based on a concrete example in bilayer QH
systems. At $\nu =3$, there are three electrons in one Landau site. We may
equivalently rephrase that there is one hole in one Landau site. Hence, we
may regard the system as a hole system at the hole filling factor $\nu
_{h}=1 $. Most discussions in the $\nu =1$ electron system go through to the 
$\nu _{h}=1$ hole system with the replacement of electrons by holes. The
symmetry breaking pattern is the same as at $\nu =1$ and there arise three
complex Goldstone modes. Solitons are $G_{4,1}$ ($CP^{3})$ solitons as in
the $\nu =1 $ case.

We proceed to discuss briefly the $N$-layer QH system, where tunneling
interactions operate between two adjacent layers. It is described by the
tunneling term consisting of an $N\times N$ matrix, 
\begin{equation}
H_{T}=-\frac{1}{2}\Delta _{\text{SAS}}\left( 
\begin{array}{cccccc}
0 & 1 & 0 & \cdots & 0 & 0 \\ 
1 & 0 & 1 & \cdots & 0 & 0 \\ 
0 & 1 & 0 & \cdots & 0 & 0 \\ 
\vdots & \vdots & \vdots & \ddots & \vdots & \vdots \\ 
0 & 0 & 0 & \cdots & 0 & 1 \\ 
0 & 0 & 0 & \cdots & 1 & 0%
\end{array}%
\right) .  \label{TunneMatri}
\end{equation}%
Let us first freeze the spin degree of freedom. By diagonalizing this matrix
the degeneracies of the $N$ energy levels are found to be resolved. The
energy of the $j$th level is%
\begin{equation}
E_{j}=\Delta _{\text{SAS}}\cos \frac{\pi j}{N+1},\quad j=1,2,\ldots ,N,
\end{equation}%
as is shown in Appendix B. The lowest energy level is unique, which gives
the ground state at $\nu =1$. At $\nu =k$ the lowest $k$ levels are
occupied. A Goldstone mode is a perturbational excitation from one of the $k$
occupied levels to one of the $N-k$ empty levels. Thus, the number of
Goldstone modes is $k(N-k)$, which is the dimension of the Grassmannian
manifold $G_{N,k}$. All Goldstone modes are made massive due to the
tunneling interactions. When the spin degree of freedom is included together
with the Zeeman interaction, each of these $N$ levels is split into two
levels by the Zeeman energy. All Goldstone modes are massive. At $\nu =k$
the lowest $k$ levels are occupied: There arise $k$ phases depending on the
relative strength between the Zeeman and tunneling energies.

\section{Discussions}

We have developed an algebraic method to explore the dynamics of electrons
in the noncommutative plane. For this purpose we have introduced the Weyl
ordering of the second quantized density operator. By making the LLL
projection we have constructed the Hamiltonian for these electrons
interacting via the Coulomb potential. It is given by (\ref{ExchaHamilMom})
in the momentum space and by (\ref{ExchaHamilX}) in the coordinate space.
The density operators make the W$_{\infty }$(N) algebra (\ref{WAlgeb}). Then
we have made a derivative expansion of the Coulomb potential and derived the
effective Hamiltonian appropriate for a description of long-distance physics
of electrons confined to the LLL. It is the SU(N) nonlinear sigma model (\ref%
{ExchaSU4}).

The SU(N) nonlinear sigma model has arisen solely from the SU(N)-invariant
Coulomb interaction (\ref{CouloHamilOri}) depending only on the total
density $\rho (\boldsymbol{x})$. Namely, a modulation of the isospin $S(\boldsymbol{x%
})$ turns out to increase the Coulomb energy by affecting the density $\rho (%
\boldsymbol{x})$. The origin of the effective Hamiltonian is the
noncommutativity (\ref{NonCommuXY}), implying the density and the isospin no
longer commute as in (\ref{WAlgeb}) when electrons are confined to the LLL.
The effective Hamiltonian (\ref{ExchaSU4}) is the leading order term of the
underlying noncommutative theory.

The effective Hamiltonian turns out to be the Grassmannian $G_{N,k}$ sigma
model at the filling factor $\nu =k$, based upon which we have explored
quantum coherence in the $N$-component QH systems. We have analyzed the
Goldstone modes and topological solitons. As is well known, the existence of
massless modes in low-dimensional spaces induces an infrared catastrophe and
unstabilize the system. In QH systems there is no such a catastrophe because
all Goldstone modes are made massive due to the Zeeman and tunneling
interactions.

It is important to investigate how these perturbational and
nonperturbational objects are represented in the original noncommutative
field theory. We would like to pursue these problems in a forthcoming paper.

It is also interesting to investigate fractional QH systems. They are mapped
to integer QH systems by way of the composite-fermion picture\cite{Jain89L}.
Topological excitations in fractional QH systems are anyons, which have
fractional electric charges and obey fractional statistics\cite{Laughlin83L}%
. Such anyons have already been observed experimentally in the monolayer $%
\nu =1/3$ QH system\cite{Picciotto97N,Saminadayar97L}. The fractional
statistics stems from the statistical transmutations specific to the low
dimensional system, and represent a deep connection between the space-time
and particle statistics. The noncommutativity is also a space-time property.
Then, topological excitations in fractional QH systems are intriguing
objects inherent to these two exotic space-time properties: They are
noncommutative anyons. The study of the noncommutative anyons may reveal
novel structures of the low dimensional noncommutative space-time. The
noncommutative gauge theory has been extensively studied in the context of
D-branes with $B$ field in the string theory. Various concepts cultivated in
D-brane analysis would be applied to noncommutative anyons and be tested
experimentally in QH systems.

\appendix

\section{Weyl-Ordered Plane Wave}

We prove the basic formulas (\ref{OmegaA}) and (\ref{OmegaOmega}) for the
Weyl-ordered plane wave. It is convenient to diagonalize the coorindate $X$, 
$X|x\rangle =x|x\rangle $. The noncommutativity (\ref{NonCommuXY}) is
represented by setting%
\begin{equation}
Y=i\ell _{B}^{2}\frac{\partial }{\partial X}.
\end{equation}%
The merit of this representation is that the simple orthonormality condition
holds within the LLL,%
\begin{equation}
\langle x^{\prime }|x\rangle =\delta (x-x^{\prime }).
\end{equation}%
Since $Y$ is a shifting operator it is easy to show%
\begin{equation}
e^{i\boldsymbol{pX}}|x\rangle =\exp [-\frac{i}{2}\ell _{B}^{2}p_{x}p_{y}]\exp %
\left[ ip_{x}x\right] |x-\ell _{B}^{2}p_{y}\rangle .
\end{equation}%
Hence, 
\begin{subequations}
\begin{align}
\langle x^{\prime }|e^{i\boldsymbol{pX}}|x\rangle & =\exp [ip_{x}x-\frac{i}{2}%
\ell _{B}^{2}p_{x}p_{y}]\delta (x-x^{\prime }-\ell _{B}^{2}p_{y})
\label{AppenX} \\
& =\exp [ip_{x}x^{\prime }+\frac{i}{2}\ell _{B}^{2}p_{x}p_{y}]\delta
(x-x^{\prime }-\ell _{B}^{2}p_{y}).  \label{AppenY}
\end{align}%
We set $x^{\prime }=x$ and integrate over it, 
\end{subequations}
\begin{equation}
\int \!dx\,\langle x|e^{i\boldsymbol{pX}}|x\rangle =\frac{2\pi }{\ell _{B}^{2}}%
\delta (\boldsymbol{p}).  \label{AppenXa}
\end{equation}%
Substituting $\sum_{n}|n\rangle \langle n|=1$, we obtain 
\begin{equation}
\sum_{n}\langle n|e^{i\boldsymbol{pX}}|n\rangle =\frac{2\pi }{\ell _{B}^{2}}%
\delta (\boldsymbol{p}),
\end{equation}%
which is (\ref{OmegaA}).

We next study%
\begin{equation}
I\equiv \int d^{2}p\,\langle m|e^{-i\boldsymbol{pX}}|n\rangle \langle i|e^{i%
\boldsymbol{pX}}|j\rangle .
\end{equation}%
Substituting $\int \!dx\,|x\rangle \langle x|=1$, we find%
\begin{align}
I\equiv & \int \!dx_{m}dx_{n}dx_{i}dx_{j}\int d^{2}p\,\langle m|x_{m}\rangle
\langle x_{n}|n\rangle \langle i|x_{i}\rangle \langle x_{j}|j\rangle  \notag
\\
& \times \langle x_{m}|e^{-i\boldsymbol{pX}}|x_{n}\rangle \langle x_{i}|e^{i%
\boldsymbol{pX}}|x_{j}\rangle .
\end{align}%
We use (\ref{AppenX}) and (\ref{AppenY}), 
\begin{align}
I& \equiv \int \!dx_{m}dx_{n}dx_{i}dx_{j}\int dp_{x}dp_{y}\,\langle
m|x_{m}\rangle \langle x_{n}|n\rangle \langle i|x_{i}\rangle \langle
x_{j}|j\rangle  \notag \\
& \times \exp \left[ ip_{x}(x_{j}-x_{m})\right] \delta (x_{m}-x_{n}+\ell
_{B}^{2}p_{y})\delta (x_{i}-x_{j}-\ell _{B}^{2}p_{y})  \notag \\
& =\frac{2\pi }{\ell _{B}^{2}}\int \!dx_{m}dx_{n}\langle m|x_{m}\rangle
\langle x_{m}|j\rangle \langle i|x_{n}\rangle \langle x_{n}|n\rangle  \notag
\\
& =\frac{2\pi }{\ell _{B}^{2}}\delta _{ni}\delta _{mj},
\end{align}%
which is (\ref{OmegaOmega}).

For the sake of completeness let us prove (\ref{OmegaA}) based on the
representation (\ref{RepreDiffe}). Here, both $X$ and $Y$ are shifting
operators, 
\begin{equation}
e^{i\boldsymbol{pX}}|x,y\rangle =e^{i\boldsymbol{px}/2}|x-\ell _{B}^{2}p_{y},y+\ell
_{B}^{2}p_{x}\rangle .
\end{equation}%
Thus,%
\begin{equation}
\langle x^{\prime },y^{\prime }|e^{i\boldsymbol{pX}}|x,y\rangle =e^{i\boldsymbol{px}%
/2}\langle x^{\prime },y^{\prime }|x-\ell _{B}^{2}p_{y},y+\ell
_{B}^{2}p_{x}\rangle _{\text{LLL}}.  \label{AppenZ}
\end{equation}%
Here, it is necessary to evaluate the scalar product within the LLL. Using
the wave function (\ref{OneBodyLLL}) we find%
\begin{align}
\langle \boldsymbol{x}|\boldsymbol{y}\rangle _{\text{LLL}}& =\sum_{n=0}^{\infty
}\langle \boldsymbol{x}|n\rangle \langle n|\boldsymbol{y}\rangle  \notag \\
& ={\frac{1}{2\pi \ell _{B}^{2}}}\exp \left( i{\frac{\boldsymbol{x}\!\wedge \!%
\boldsymbol{y}}{2\ell _{B}^{2}}}\right) \exp \left( -{\frac{|\boldsymbol{x}-\boldsymbol{y%
}|^{2}}{4\ell _{B}^{2}}}\right) .
\end{align}%
From these we derive (\ref{AppenXa}) or (\ref{OmegaA}). We can prove (\ref%
{OmegaOmega}) also in this representation though slightly complicated.

\section{Tunneling Matrix\newline
}

We diagonalize the tunneling matrix (\ref{TunneMatri}), by solving the
secular equation%
\begin{equation*}
\det (H_{\text{T}}-\lambda I)=0.
\end{equation*}%
Here $H_{\text{T}}$ is given by (\ref{TunneMatri}) and $I$ is the $N\times N$
identity matrix. This equation leads to $D_{N}(x)=0$, where%
\begin{equation}
D_{N}(x)=\det \left( 
\begin{array}{cccccc}
x & 1 & 0 & \cdots & 0 & 0 \\ 
1 & x & 1 & \cdots & 0 & 0 \\ 
0 & 1 & x & \cdots & 0 & 0 \\ 
\vdots & \vdots & \vdots & \ddots & \vdots & \vdots \\ 
0 & 0 & 0 & \cdots & x & 1 \\ 
0 & 0 & 0 & \cdots & 1 & x%
\end{array}%
\right) ,
\end{equation}%
with $x=2\lambda /\Delta _{\text{SAS}}$. Expanding it with respect to the
first column we obtain the recurrence relation,%
\begin{equation}
D_{N}(x)=xD_{N-1}(x)-D_{N-2}(x).
\end{equation}%
Evidently, $D_{1}(x)=x$ and $D_{2}(x)=x^{2}-1$, and we can define $%
D_{0}(x)=1 $ from this relation.

It is crucial to recall that the Chebyshev polynomial $S_{N}(x)$ satisfies
the same recurrence relation (see 22.7.6 in Ref.\cite{Abramowitz}). We use
22.5.13 and 22.5.48 in this reference to find that%
\begin{equation}
S_{N}(x)=U_{N}\bigg(\frac{x}{2}\bigg)=(N+1)F\bigg(-N,N+2;\frac{3}{2};\frac{%
2-x}{4}\bigg),
\end{equation}%
where $F(a,b;c;z)$ is the hypergeometric function. Since $a=-N$, the
hypergeometric function becomes truncated. We can easily check that $%
S_{0}(x)=1$ and $S_{1}(x)=x$.

Therefore, we may identify%
\begin{equation}
D_{N}(x)=U_{N}\bigg(\frac{x}{2}\bigg)=U_{N}\bigg(\frac{\lambda }{\Delta _{%
\text{SAS}}}\bigg).
\end{equation}%
According to 22.16.5 in the same reference we get the following set of roots,%
\begin{equation}
\lambda _{j}(N)=\Delta _{\text{SAS}}\,\mathrm{\cos }\frac{\pi j}{N+1},\quad
j=1,2,\ldots ,N.
\end{equation}%
We note that $\lambda _{j}=-\lambda _{N-j+1}$. For $N=2K$ the roots with $%
j=1,\ldots ,K$ are positive. The ones with $j=K+1,\ldots ,2K$ are of the
same magnitude but negative. We have no zero root. For $N=2K+1$ there is one
zero root corresponding to $j=K+1$, the ones with $j=1,\ldots ,K$ are
positive, while those with $j=K+2,\ldots ,2K$ are negative with the same
magnitude. The lowest root is%
\begin{equation}
\lambda _{N}=-\Delta _{SAS}\mathrm{\cos }\frac{\pi }{N+1},
\end{equation}%
and the next root is%
\begin{equation}
\lambda _{N-1}=-\Delta _{SAS}\mathrm{\cos }\frac{2\pi }{N+1}.
\end{equation}%
There is no degeneracy among $N$ energy levels.

\section*{Acknowledgements}

We would like to thank Merab Eliashvili and Jnanadeva Maharana for fruitful
discussions on the subject. One of the authors (ZFE) is grateful to the
hospitality of Theoretical Physics Laboratory, RIKEN, where a part of this
work was done. Another author (GT) expresses his sincere gratitude to
Nishina Memorial Foundation for supporting his stay at Department of
Physics, Tohoku University, where a part of this work was carried out. ZFE
is supported in part by Grants-in-Aid for Scientific Research from Ministry
of Education, Science, Sports and Culture (Nos.13135202,14540237); GT by the
grant\ from SCOPES (No.7GEPJ62379); KH by the grant from the
Interdisciplinary Research Program of the KOSEF (No.R01-1999-00018) and the
special grant of Sogang University in 2002.

\end{document}